\documentclass[useAMS,usenatbib]{mn2e}

\pdfoutput=1
\usepackage{graphicx}
\usepackage{color}
\usepackage{latexsym}
\usepackage{rotating}
\usepackage{lscape}
\usepackage{dcolumn}
\usepackage{amsmath}
\usepackage{amssymb}
\usepackage{textcomp}

\title[Multi-molecular lines mapping of the GMC associated with RCW 106]
{Molecular line mapping of the giant molecular cloud associated with RCW 106 - {\LARGE III.
Multi-molecular line mapping}}

\author[N. Lo et al.]{N.~Lo,$^{1,2}$\thanks{E-mail: nlo@phys.unsw.edu.au}
M. R.~Cunningham,$^{1}$ P. A.~Jones,$^{1,3}$ I.~Bains,$^{1,4}$ M. G.~Burton,$^{1}$ \and
T.~Wong,$^{1,2,5}$ E.~Muller,$^{2,6}$\thanks{Bolton Fellow, ATNF} C.~Kramer,$^{7}$
V.~Ossenkopf,$^{8,9}$ C.~Henkel,$^{10}$ \and
G.~Deragopian,$^{11}$ S.~Donnelly$^{1}$ and E. F.~Ladd$^{12}$\\
\\
$^{1}$School of Physics, University of New South Wales, Sydney, NSW 2052, Australia\\
$^{2}$Australia Telescope National Facility, CSIRO, PO Box 76, Epping, NSW 1710, Australia\\
$^{3}$Departamento de Astronom\~Aa, Universidad de Chile, Casilla 36-D, Santiago, Chile \\
$^{4}$Centre for Astrophysics and Supercomputing, Swinburne University of Technology, P.O.
 Box 218, Hawthorn, VIC 3122, Australia\\
$^{5}$Astronomy Department, University of Illinois, 1002 W. Green St, Urbana, IL 61801, USA\\
$^{6}$Department of Astrophysics, Nagoya University, Furo-cho, Chikusa-ku, Nagoya 464-8602,
 Japan\\
$^{7}$Instituto de Radioastronomia Milimetrica (IRAM), Avda. Divina Pastora 7, E-18012 Granada, Spain\\
$^{8}$I. Physikalisches Institut, Universit{\"a}t zu K{\"o}ln, Z{\"u}lpicher Stra$\beta$e 77, 50937 K{\"o}ln, Germany\\
$^{9}$SRON, Netherlands Institute for Space Research, PO Box 800, 9700 AV Groningen, The
 Netherlands\\
$^{10}$Max-Planck-Institut f{\"u}r Radioastronomie, Auf dem H$\ddot{u}$gel 69, 53121 Bonn,
 Germany\\
$^{11}$Centre for Astronomy, James Cook University, Townsville, Australia\\
$^{12}$Department of Physics and Astronomy, Bucknell University, Lewisburg PA 17837, USA}

\begin{document}

\date{Accepted ***. Received ***; in original form ***}

\pagerange{\pageref{firstpage}--\pageref{lastpage}} \pubyear{}

\maketitle

\label{firstpage}

\begin{abstract}
We present multi-molecular line maps obtained with the Mopra Telescope towards the southern giant molecular cloud (GMC) complex G333, associated with the H{\sc ii} region RCW 106. We have characterised the GMC by decomposing the 3D data cubes with {\sc gaussclumps}, and investigated spatial correlations among different molecules with principal component analysis (PCA). We find no correlation between clump size and line width, but a strong correlation between emission luminosity and line width. PCA classifies molecules into high and low density tracers, and reveals that HCO$^+$ and N$_2$H$^+$ are anti-correlated.
\end{abstract}


\section{Introduction}
This is the third paper (paper I by \citeauthor{Bains2006} \citeyear{Bains2006} and paper II by \citeauthor{Wong2008} \citeyear{Wong2008}) in a series of multi-molecular lines observations of the giant molecular cloud (GMC) associated with RCW 106, G333. The G333 giant molecular cloud, a southern massive star forming region, spans $1\fdg2 \times 0\fdg6$ on the sky, roughly centred on \textit{l} $\sim 333^{\circ}$, \textit{b} $\sim -0\fdg5$ ($\alpha_{J2000} = 16^h21^m, \delta_{J2000} = -50^d30^m$), at a distance of 3.6 kpc \citep{Lockman1979}. The aim of the multi-molecular line study was to investigate the relationship between the dynamics of the interstellar medium and star formation. Further investigation and analysis planned include power spectra to study the role of turbulence in the GMC, and searching for evidence of triggered star formation. \\
\indent The G333 GMC has previously been studied in other wavelengths, for example in the far-infrared by \citet{Karnik2001} and the 1.2-mm dust continuum by \citet{Mookerjea2004}. A search for water masers was carried out by \citet{Breen2007}. There are also numerous works on specific regions in this GMC, such as RCW 106 (e.g. \citeauthor*{Rodgers1960} \citeyear{Rodgers1960}, \citeauthor{Russeil2005} \citeyear{Russeil2005}) and G333.6$-$0.2, one of the brightest southern compact H{\sc ii} regions (e.g. \citeauthor*{Goss1970} \citeyear{Goss1970}, \citeauthor{Becklin1973} \citeyear{Becklin1973}, \citeauthor{Storey1989} \citeyear{Storey1989}, \citeauthor{Colgan1993} \citeyear{Colgan1993}, \citeauthor{Fujiyoshi1998} \citeyear{Fujiyoshi1998}, \citeyear{Fujiyoshi2001}, \citeyear{Fujiyoshi2005} and \citeyear{Fujiyoshi2006}). However, to date, there has been no systematic, 3-mm multi-molecular line mapping of this GMC. This study demonstrates the full capability of the Mopra Telescope's\footnote{The Australia Telescope Mopra telescope is part of the Australia Telescope, which is funded by the Commonwealth of Australia for operation as a National Facility managed by CSIRO (ATNF, http://www.atnf.csiro.au).} new digital filter bank, the UNSW-Mopra Spectrometer (UNSW-MOPS). Other similar surveys utilising the UNSW-MOPS are the H$_2$O Southern Galactic Plane Survey (HOPS) by \citet{Walsh2008} and the Central Molecular Zone of the Galaxy by \citet{Jones2008}.\\
\indent We present in this paper results of molecular line mapping carried out during July to November in 2006, consisting of molecular lines from 83 to 101 GHz. For a detailed analysis of $^{13}$CO and C$^{18}$O data please refer to previous work by \citeauthor{Bains2006} (\citeyear{Bains2006}, hereafter BWC2006) and \citeauthor{Wong2008} (\citeyear{Wong2008}, hereafter WLB2008). In Section \ref{sec:obs} we describe the observing technique, and the new UNSW-MOPS digital filter bank. In Section \ref{sec:results} we present the results, and examine the velocity and spatial distribution of the molecular emission from the different molecules. In Section \ref{sec:clumps} and \ref{sec:PCA}, we present two different approaches to characterising emission distribution in the GMC: Clump finding with {\sc gaussclumps} and principal component analysis (PCA). In Section \ref{sec:discussion} we discuss results, including specific regions of the GMC. Finally, we summarise our findings in Section \ref{sec:summary}.\\
\indent Data presented here will be made available in the near future, please contact the authors for details.\\
\begin{table*}
  \caption{List of observed molecular transitions mapped in G333.
  The columns are: (1) molecule; (2) transition; (3) rest frequency 
  \citep*{Lovas1979}; (4) upper energy levels from the Cologne Database for
  Molecular Spectroscopy \citep[CDMS,][]{{Muller2001},{Muller2005}};
  (5) calculated critical density: unless specified, $n_{crit}=A_{ul}/\langle\sigma(v)v\rangle$
  where $A_{ul}$ is the Einstein A coefficient obtained from CDMS,
  $\langle\sigma(v)\rangle=10^{-15}$ cm$^2$ is
  the collision cross section and $v$ is the velocity of collision particles assumed to be 1
  km s$^{-1}$; (6) observation season; (7) backend used, where MPCOR stands for Mopra
  Correlator, which was decommissioned in 2005 and replaced with a new digital filter bank,
  the UNSW-Mopra Spectrometer (UNSW-MOPS); (8) notes, where [hf] denotes hyperfine structures,
  with the rest frequency quoted being that of the strongest hyperfine component, and
  [\dag] denotes molecules that have detectable emission at a few places only. Their emission
  maps are not presented in this work but will appear in subsequent papers on specific regions in 
  the G333 cloud.}
  \label{tab:mol_list}
  \newcolumntype{.}{D{.}{.}{-1}}
  \begin{minipage}{1\textwidth}
  \centering
  \begin{tabular}{c c . c c c c c c}
  \hline
  Molecule & Transition & \multicolumn{1}{c}{Rest frequency} & $E_u/k$ & $n_{crit}$
   & Observation season & Backend & Notes \\
   & & \multicolumn{1}{c}{\scriptsize (GHz)} & \scriptsize (K) & \scriptsize (cm$^{-3}$) & & & \\
  \hline \hline
  $^{13}$CO & $1-0$ & 110201.353 & 5.29 & $3\times10^3$ & 2004 & MPCOR & \\
  C$^{18}$O & $1-0$ & 109782.173 & 5.27 & $3\times10^3$ & Jul, 2005 & MPCOR & \\
  C$_2$H & $1-0$ & 87316.925 & 4.19 & $2\times10^5$ & Jul, 2006 & MOPS & [hf] \\
  CH$_3$OH & $2(0,2)-1(0,1)\, A+$ & 96741.377 & 6.97 & $7\times10^3$ & Sep, 2006 & MOPS
    & [\dag] \\
  CS & $2-1$ & 97980.953 & 7.05 & $2\times10^5$ & Sep, 2006 & MOPS & \\
  C$^{34}$S & $2-1$ & 96412.961 & 6.94 & $2\times10^5$ & Sep, 2006 & MOPS & [\dag] \\
  HCN & $1-0$ & 88631.847 & 4.25 & $2\times10^5$ & Jul, 2006 & MOPS & [hf]\\
  H$^{13}$CN & $1-0$ & 86340.167 & 4.14 & $2\times10^5$ & Jul, 2006 & MOPS & [hf,\dag] \\
  HCCCN & $10-9$ & 90978.989 & 24.01 & $6\times10^5$ & Sep, 2006 & MOPS & [\dag] \\
  HCCCN & $11-10$ & 100076.385 & 28.82 & $8\times10^5$ & Sep, 2006 & MOPS & [\dag] \\
  HCO$^+$ & $1-0$ & 89188.526 & 4.28 & $4\times10^5$ & Jul, 2006 & MOPS & \\
  H$^{13}$CO$^+$ & $1-0$ & 86754.330 & 4.16 & $4\times10^5$ & Jul, 2006 & MOPS & [\dag] \\
  HNC & $1-0$ & 90663.572 & 4.35 & $3\times10^5$ & Jul, 2006 & MOPS & [hf] \\
  N$_2$H$^+$ & $1-0$ & 93173.480 & 4.47 & $4\times10^5$ & Sep, 2006 & MOPS & [hf] \\
  SiO & $2-1\,\nu=0$ & 86847.010 & 6.25 & $3\times10^5$ & Jul, 2006 & MOPS & [\dag] \\
  SO & $3(2)-2(1)$ & 99299.905 & 9.23 & $1\times10^5$ & Sep, 2006 & MOPS & [\dag] \\
   & $4(5)-4(4)$ & 100029.565 & 38.58 & $3\times10^7$\footnote[1]{\citet{Ungerechts1997}}
     & Sep, 2006 & MOPS & [\dag] \\
  \hline
  \end{tabular}
  \end{minipage}
\end{table*}
\section{Observations}\label{sec:obs}
The data were collected with the 22 metre Mopra Telescope, which is a centimetre- and millimetre-wavelength antenna having a full width to half-maximum (FWHM) beam size of $\sim$36\arcsec at 100-GHz \citep{Ladd2005}. The observations were carried out with the narrow band mode of the new UNSW-Mopra Spectrometer (UNSW-MOPS) digital filterbank back-end, and a Monolithic Microwave Integrated Circuit (MMIC) 77 to 116 GHz receiver. The observing parameters, including rest frequencies and dates are listed in Table \ref{tab:mol_list}. UNSW-MOPS has a 8-GHz bandwidth with four overlapping 2.2-GHz subbands, each subband having four dual-polarisation 137.5-MHz-wide windows giving a total of sixteen dual-polarisation windows. Each window has 4096 channels providing a velocity resolution of $\sim$0.1 km s$^{-1}$ per channel at 100 GHz. The velocities presented in this work are with respect to the kinematic local standard of rest (LSR), with $v_{\,\rm{LSR}} = -50$ km s$^{-1}$ being the systemic velocity of the G333 complex. \\
\indent The brightness temperature $T_\mathrm{b}$ is related to the antenna temperature $T^*_\mathrm{A}$ by $T_\mathrm{b} = T^*_\mathrm{A} / \eta_\nu$, where $\eta_\nu$ is the frequency dependent beam efficiency. According to \citet{Ladd2005} the main beam efficiency at 86 GHz is $\eta_\mathrm{\,86\,GHz} = 0.49$, and at 110 GHz is $\eta_\mathrm{\,110\,GHz} = 0.44$. The beam efficiencies are $\eta_\mathrm{\,86\,GHz} = 0.65$ and $\eta_\mathrm{\,86\,GHz} = 0.56$ at 86 and 110 GHz respectively. The results presented in this paper are in terms of antenna temperature $T^*_\mathrm{A}$ unless otherwise specified.\\
\indent We followed a similar mapping procedure to our previous work ($^{13}$CO; BWC2006).
The $\sim$1 square degree region was divided into 300 $\times$ 300 square arcsec fields, with pointing centres separated by 285 arcsec in right ascension and declination, allowing a 15-arcsec overlap between adjacent fields. Each field was mapped with two passes, the first pass scanning in right ascension and the second pass scanning in declination. The observing mode was `on-the-fly'(OTF) raster scanning, at a scan rate of 3.5 arcsec s$^{-1}$ and averaging data over a 2-s cycle time. We used the $^{13}$CO map as a guide, mapping the brightest $^{13}$CO fields first, then extending these to cover the majority of the emission in most detected molecules, resulting in maps covering $\sim$0.7 square degrees. We used two frequency settings centred at 87 and 97 GHz, with each setting covering an 8-GHz bandwidth, 4-GHz on either side of the central frequency.\\
\indent We have used the SiO masers IRSV1540 and AH Scorpii for pointing calibration. Pointing calibration was performed at the beginning of each map, therefore approximately every 50 minutes. Estimated pointing errors are within 10 arcseconds. \\
\indent At the end of each day's observation, a 5 minute position switched observation of M17SW was carried out with the same frequency settings as for the maps, to monitor the performance of the system. Based on these observations the deviation of T$_A^*$ during the whole observing season is within 10 per cent.\\
\indent The data were reduced with the {\sc livedata} and {\sc gridzilla} packages available from the ATNF, written by Mark Calabretta\footnote{http://www.atnf.csiro.au/computing/software/}. {\sc livedata} performs a bandpass calibration using the preceding reference (off-source) scan, then fits the spectral baseline with a first degree polynomial. Due to low sensitivity near the start and end of the window, the first and last 750 channels were discarded before baseline fitting. {\sc gridzilla} grids the data according to user specified weighting and beam parameter inputs. In this work, our data was weighted by the relevant system temperature (T$_{sys}$) and Gaussian smoothed during data reduction stage. The data cubes were gridded with three pixels across each beam width, therefore each pixel is $\sim$ 15 $\times$ 15 arcsec$^2$ for data with 87-GHz frequency setting, and $\sim$ 12 $\times$ 12 arcsec$^2$ for the 97-GHz setting. Note that all data cubes presented in this work are regridded to 15 $\times$ 15 arcsec$^2$ pixel size for ease of comparison, and the velocity resolution is left as $\sim0.1$ km s$^{-1}$.\\

% ########################################################
%   Results
% ########################################################
\section{Results}\label{sec:results}
%%% spatial distributions %%%
\subsection{Spatial distribution of the molecules}\label{sec:maps}
\begin{figure*}
  \centering
  \includegraphics[]{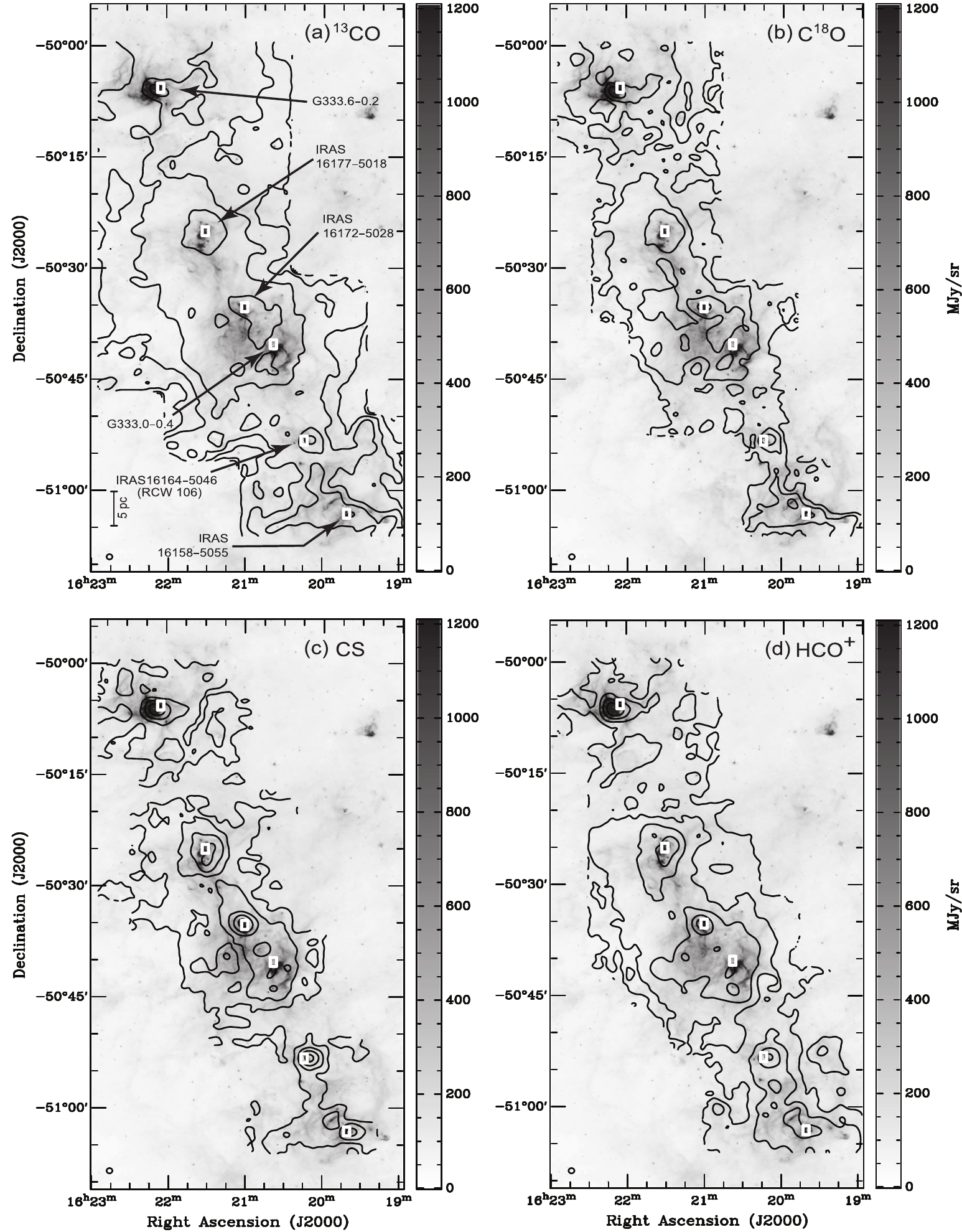}
  \caption[]{Integrated emission maps (contours) for $^{13}$CO ($J=1-0$), C$^{18}$O ($J=1-0$), CS
  ($J=2-1$) and HCO$^+$ ($J=1-0$) overlaid on a {\it Spitzer} IRAC 8.0-$\mu m$ image (grey scale). 
  The maps are integrated over a velocity range of $-70$ to $-40$ km s$^{-1}$ and were clipped at a 3 
  $\sigma$ level. 1 $\sigma$ is 0.3 K km s$^{-1}$ for $^{13}$CO,
  0.2 K km s$^{-1}$ for C$^{18}$O and CS, and 0.1 K km s$^{-1}$ for HCO$^+$. The maps were 
  then Gaussian smoothed with an FWHM of 45 arcsec. 
  The contour levels start at 1.5 K km s$^{-1}$ for $^{13}$CO, 2.0 K km s$^{-1}$ for
  C$^{18}$O, 1.4 K km s$^{-1}$ for CS and 1.0 K km s$^{-1}$ for HCO$^+$. Then each successive contour level 
  is double the value of the previous one. The choice of lowest contour level is based on the
  lowest detectable emission above noise, except for $^{13}$CO where there is emission detected throughout the map. 
  The temperatures are in terms of the antenna temperature, $T^*_{\rm A}$. 
  Previously designated H{\sc ii} regions and the associated {\it IRAS} sources are marked with 
  white rectangles and names are shown in the $^{13}$CO map. 
  The scale bar shown in the $^{13}$CO map indicates 5-pc at a
  distance of 3.6-kpc and the tiny circle in the lower left corner of each panel show the FWHM beam size after smoothing. 
  The velocity range used here
  reflects emission from the GMC only; emission outside this range comes from more distant spiral 
  arms or intervening clouds.}
  \label{fig:int1}
\end{figure*}

\begin{figure*}
  \centering
  \includegraphics[]{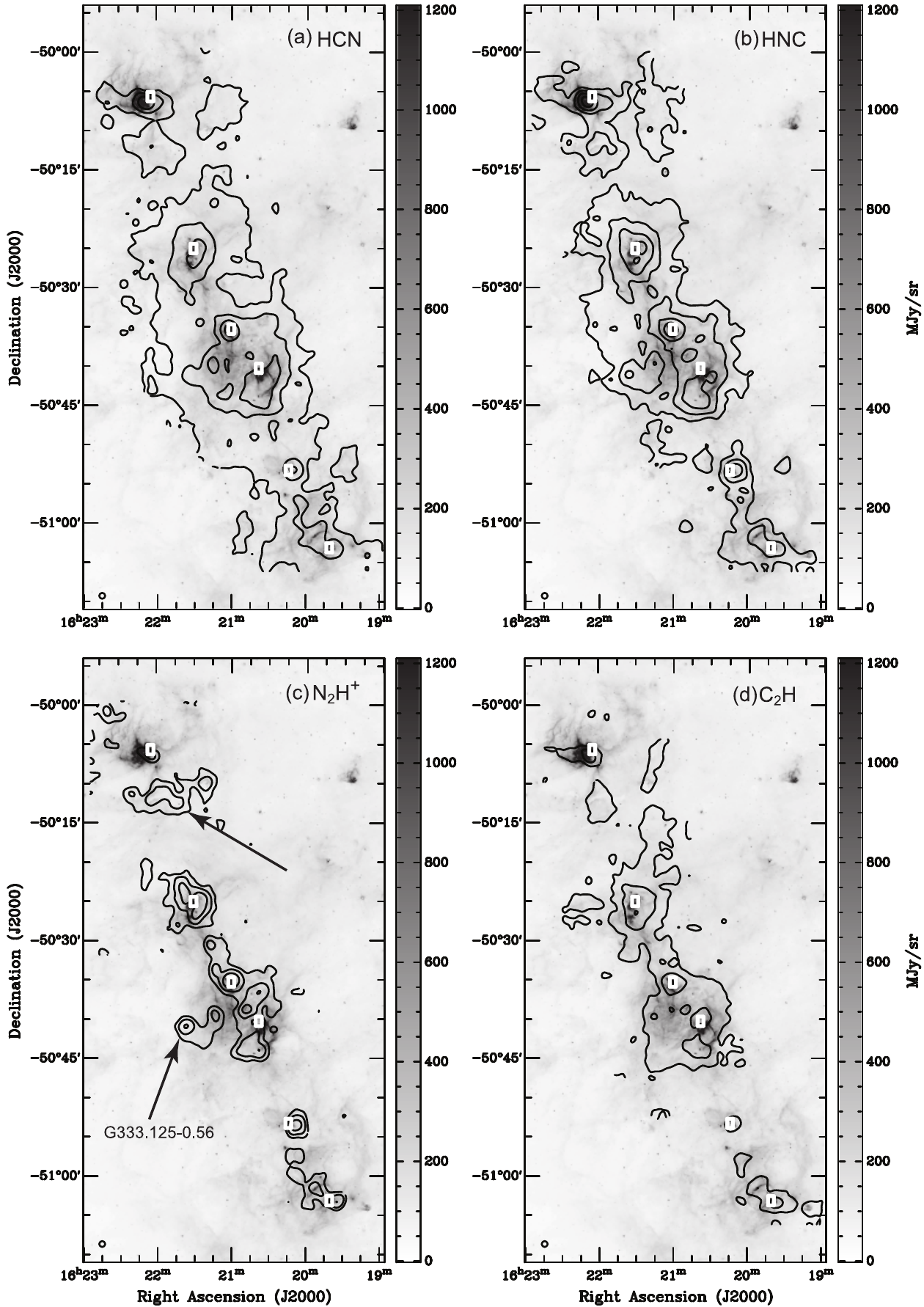}
  \caption[]{Integrated emission maps (contours) for HCN ($J=1-0$), HNC ($J=1-0$), N$_2$H$^+$
  ($J=1-0$) and C$_2$H ($J=1-0$) overlaid on \textit{Spitzer} IRAC 8-$\mu m$ images (grey scale).
  The integrated velocity ranges are of $-80$ to $-30$ km s$^{-1}$ for
  HCN and N$_2$H$^+$, $-70$ to $-40$ km s$^{-1}$ for HNC and $-60$ to $-40$ km s$^{-1}$
  for C$_2$H.
  The maps were clipped at a 3 $\sigma$ level. 1 $\sigma$ is 0.1 K km s$^{-1}$ for HCN and
  HNC, and 0.2 K km s$^{-1}$ for N$_2$H$^+$ and C$_2$H. The maps were then Gaussian smoothed 
  with an FWHM of 45 arcsec. 
  The contour levels start at 2.0 K km s$^{-1}$ for HCN, 1.3 K km s$^{-1}$ for HNC, 1.6 K km s$^{-1}$ 
  for N$_2$H$^+$ and 1.2 K km s$^{-1}$ for C$_2$H, then each successive level is double the
  previous one. The choice of lowest contour levels are based on the
  lowest detectable emission above noise.
  The temperatures are in terms of $T^*_{\rm A}$ and the tiny circle in the lower left corner of each panel show the FWHM beam 
  size after smoothing. 
  Previously designated H{\sc ii} regions and the associated {\it IRAS} sources are marked with 
  white rectangles (see Figure \ref{fig:int1} for their naming).
  The arrows indicate two N$_2$H$^+$ emission peaks, which are not prominent
  in other emission maps. Note that the
  differences in velocity range between molecules are due to hyperfine splitting such
  that some molecules have a wider velocity distribution. The velocity range used here
  reflects emission from the GMC only; emission outside this range is from more distant spiral 
  arms or intervening clouds.}
  \label{fig:int2}
\end{figure*}

Different molecules may trace different physical conditions within the region, such as density, chemistry, temperature, evolutionary stage, and so on. Our selection of molecules covers those which may trace differences in density, temperature and dynamic features such as outflow, allowing us to study environmental varieties within this giant molecular cloud. Since some molecules (e.g. CH$_3$OH, HCCCN) have detectable emission at a few places only, their emission maps and velocity profiles are not presented here, but in subsequent works focusing on specific regions. \\
\indent Shown in Figure \ref{fig:int1} and \ref{fig:int2} are the integrated intensity (zeroth moment) maps of the observed molecular lines (contours) overlaid on the \textit{Spitzer} Galactic Legacy Infrared Midplane Survey Extraordinaire \citep[GLIMPSE; ][]{Benjamin2003} 8.0-$\mu m$ image (grey scale). The integrated velocity range is $-70$ to $-40$ km s$^{-1}$ for $^{13}$CO (BWC2006), C$^{18}$O (WLB2008), CS, HCO$^+$ and HNC, and $-60$ to $-40$ km s$^{-1}$ for C$_2$H. A wider velocity range was chosen for molecules with hyperfine components: $-80$ to $-30$ km s$^{-1}$ for HCN and N$_2$H$^+$, so as to include emission from all components. The choice of velocity range reflects emission from the GMC only. Emission outside this range comes from spiral arms and intermittent clouds. The maps were Gaussian smoothed with an FWHM of 45 arcsec. The labels in the $^{13}$CO total intensity map (Figure \ref{fig:int1}a) indicate some of the molecular regions in proximity to bright H{\sc ii} regions and {\it IRAS} sources, as marked by rectangles. The starting contour levels are based on the lowest detectable emission above noise, hence outlining the emission structure of the cloud. Note that contours near the edges of the maps are noisy; this is because the edges were scanned with one pass only and thus have lower sensitivity. Details of the integrated emission maps are summarised in Table \ref{tab:map_detail}.\\
\begin{table}
  \centering
  \caption{A list of the emission range, peak brightness and 1$\sigma$ level of the integrated
    emission maps shown in Figures \ref{fig:int1} and \ref{fig:int2}.}
  \label{tab:map_detail}
  \begin{tabular}{c c c c}
    \hline
    Molecule & Emission range & 1 $\sigma$ & Peak brightness \\
     & \scriptsize (km s$^{-1}$) & \scriptsize (K km s$^{-1}$) & \scriptsize (K km s$^{-1}$) \\
    \hline \hline
    $^{13}$CO & $-70$ to $-40$ & 0.3 & 95 \\
    C$^{18}$O & $-70$ to $-40$ & 0.2 & 19 \\
    CS & $-70$ to $-40$ & 0.2 & 30 \\
    HCO$^+$ & $-70$ to $-40$ & 0.1 & 21 \\
    HCN & $-80$ to $-30$ & 0.1 & 21 \\
    HNC & $-70$ to $-40$ & 0.1 & 18 \\
    N$_2$H$^+$ & $-80$ to $-30$ & 0.2 & 11 \\
    C$_2$H & $-60$ to $-40$ & 0.2 & 5 \\
    \hline
  \end{tabular}
\end{table}
\indent The CS total intensity map (Figure \ref{fig:int1}c) shows a more confined distribution than the $^{13}$CO emission. This is as expected from a dense gas tracer such as CS, with a critical density of $\sim10^5$ cm$^{-3}$. However CS has a similar distribution to another CO isotope presented here $-$ C$^{18}$O. HCO$^+$ (Figure \ref{fig:int1}d), a common ionic species found in molecular clouds also presents a similar distribution to CS, as do HCN (Figure \ref{fig:int2}a) and HNC (Figure \ref{fig:int2}b).\\
\indent A comparison of HCN and HNC total intensity maps (Figures \ref{fig:int2}a and b), shows that they both have similar distributions in general. This is surprising because of the strong chemical differences between HCN and HNC, meaning that its abundance ratio can depend heavily on, for example, temperature \citep[e.g.][]{{Schilke1992},{Hirota1998}}. We do note there are local variations where temperature changes, such as the massive dense cold core where HCN and HNC have different spatial distribution \citep{Lo2007}. \\
\indent The total emission map of N$_2$H$^+$ (Figure \ref{fig:int2}c) shows that this molecule has a more compact and clumpier distribution than other molecules. It also shows regions with intense N$_2$H$^+$ emission which show diffuse emission in other molecules; here we highlight two such regions on the map (indicated by arrows). The source G333.125$-$0.56 (bottom arrow) is a massive dense cold core, with temperatures of 13 to 19 K derived from the NH$_3$ (J,K) = (1,1) and (2,2) inversion line and from the spectral energy distribution (SED). It has broad thermal SiO emission, and is believed to be a deeply embedded young stellar object in an early stage of star formation \citep{Lo2007}. In the northern part of the map, the N$_2$H$^+$ emission indicated by the top arrow is similarly interesting. Here, bright N$_2$H$^+$ emission is detected where other molecules show diffuse weak emission. A comparison with the GLIMPSE 8.0-$\mu$m image clearly shows that N$_2$H$^+$ aligns well with the infrared dark filaments, as discussed later in Section \ref{sec:IRring}. While N$_2$H$^+$ seems to be correlated with infrared dark filaments, C$_2$H does the opposite; it seems to correlate with GLIMPSE 8.0-$\mu$m emission. \\
\indent Among the eight molecules we present, C$_2$H emission is the weakest. Its spatial distribution is more extended than the quiescent gas tracer N$_2$H$^+$, but not as extended as other molecules. \\

%%% averaged spectra
\subsection{Velocity distribution of the molecules}\label{sec:vel_dist}
\begin{figure*}
  \centering
  \includegraphics[]{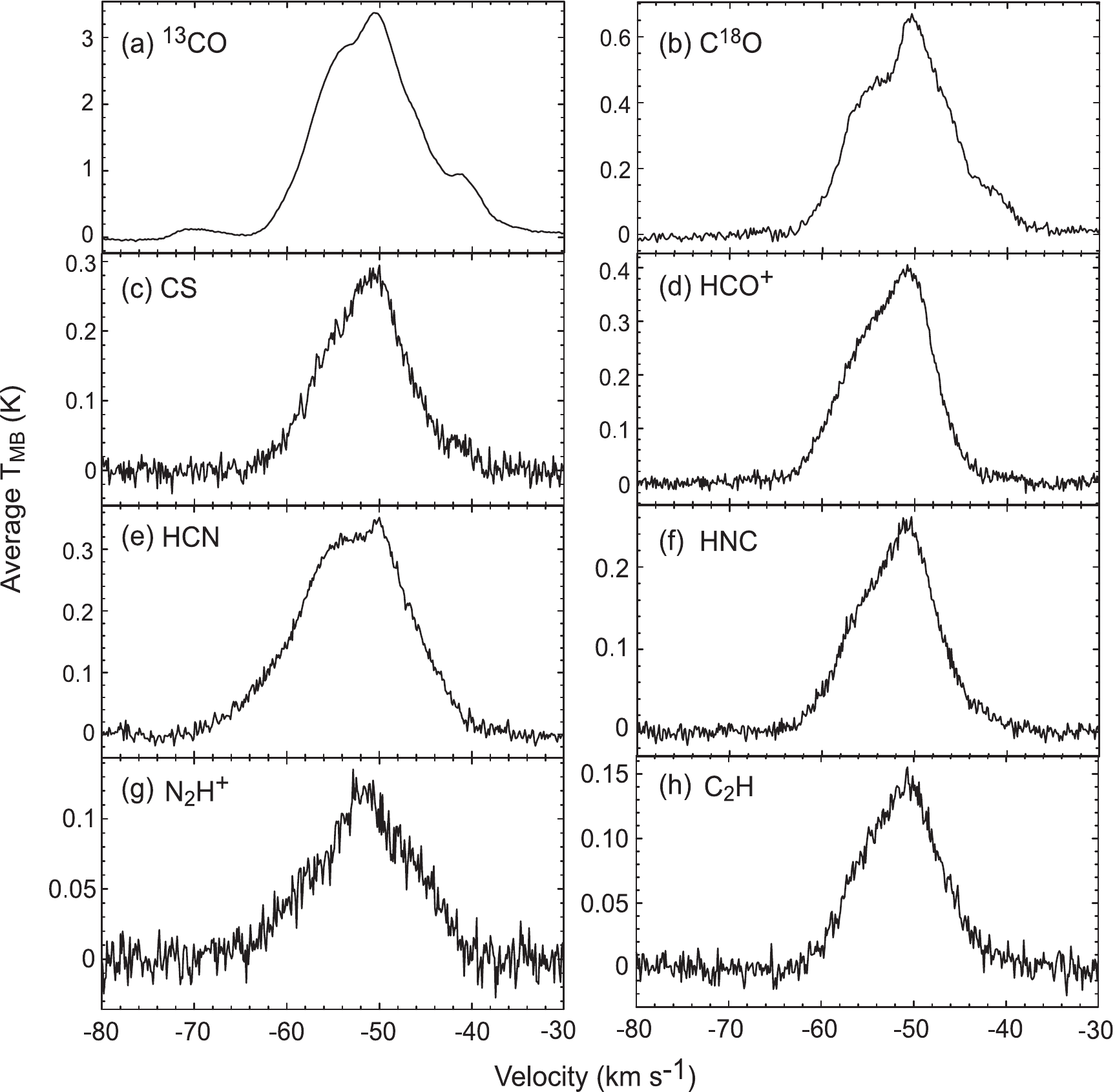}
  \caption[]{Spatially averaged emission profiles of the whole GMC. The temperatures are in
  terms of $T^*_{\rm A}$.
  Note that the lower signal-to-noise ratio of the N$_2$H$^+$ and C$_2$H spectra is due to these
  two molecules being less extended spatially; the 1 $\sigma$ level is comparable to other more 
  extended molecules.}
  \label{fig:avespec}
\end{figure*}
\begin{figure*}
  \centering
  \includegraphics[]{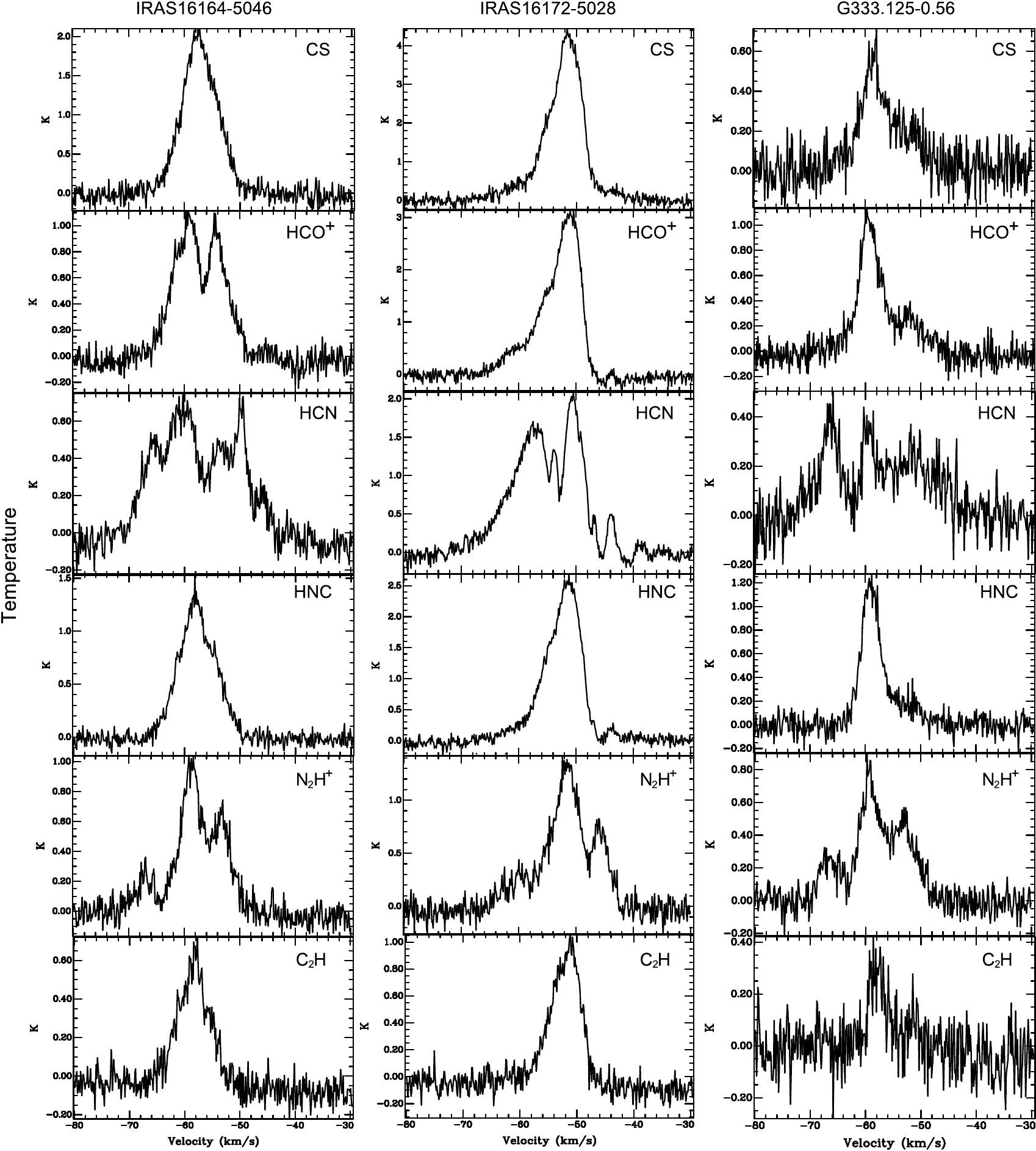}
  \caption[]{Sample spectra towards two of the IRAS sources (IRAS16172$-$5028 and IRAS16164$-$5046) and the cold core G333.125$-$0.56 in G333, showing the general quality of the data, and the consistent noise level within the data set. Line temperatures are given on a $T^*_{\rm A}$ scale.}
  \label{fig:region_spectra}
\end{figure*}
\begin{figure*}
  \centering
  \includegraphics[]{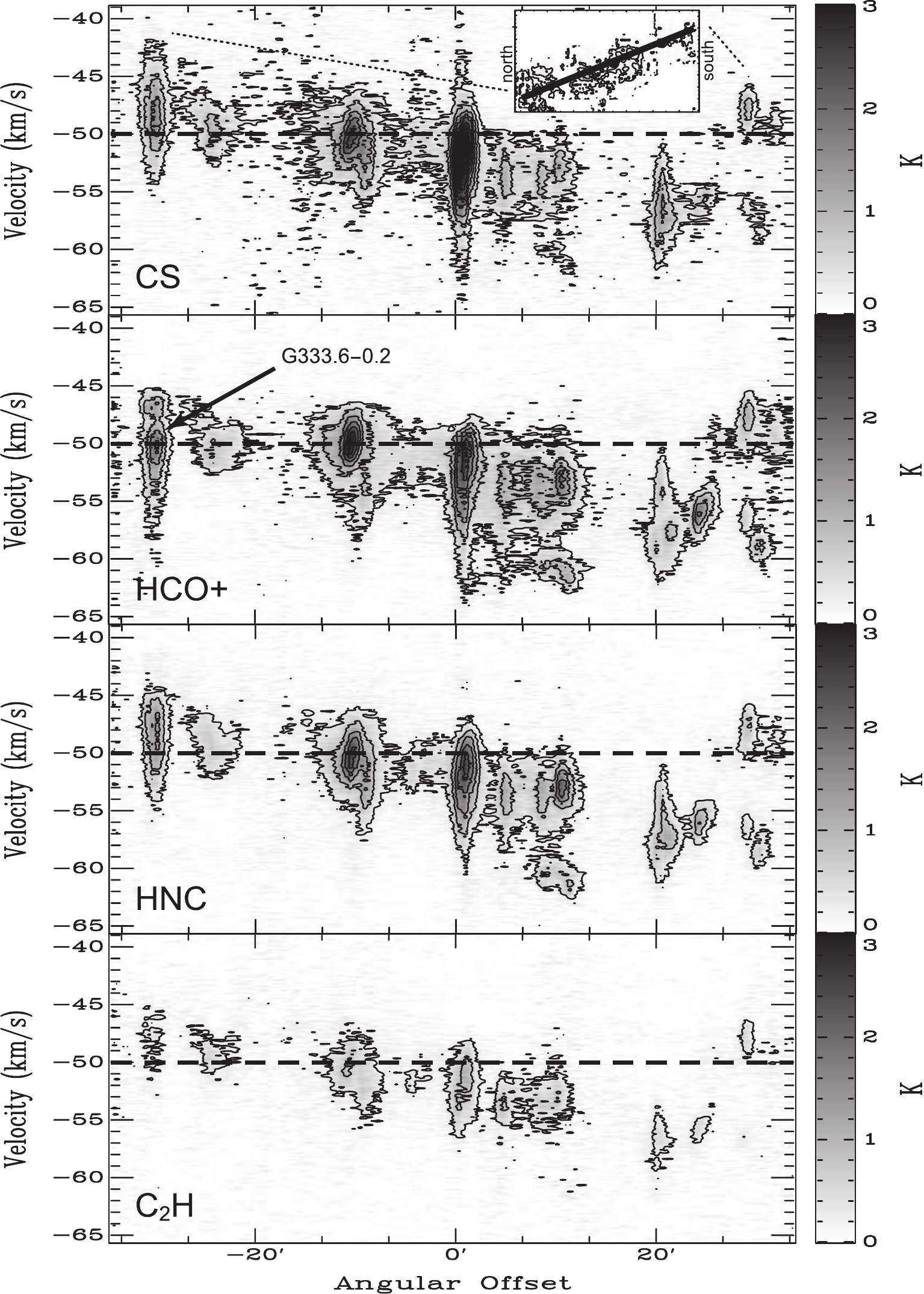}
  \caption[]{Position-velocity diagrams of CS, HCO$^+$, HNC and C$_2$H centred at
  approximately $\alpha_{J2000} = 16^h21^m$, $\delta_{J2000} = -50^d$30$^m$. The inset in the top panel
  shows the position-velocity cut presented in the diagrams. The contour
  levels all start at the 3$\sigma$ level, 0.3 K, increasing to 3.0 K with increments of 0.5 K. The 
  temperatures are in terms of $T^*_{\rm A}$.
  Positive angular offset corresponds to the southern region of the GMC. The arrow points toward the H{\sc
  ii} region G333.6$-$0.2.}
  \label{fig:pv_map}
\end{figure*}

Spatially averaged spectral emission profiles of the molecules presented in Section \ref{sec:maps} are shown in Figure \ref{fig:avespec}. Note that the apparent lower signal-to-noise of the N$_2$H$^+$ and C$_2$H spectra is due to these two molecules being less extended spatially. To illustrate this, spectra of specific regions (IRAS16172$-$5028, IRAS16164$-$5046 and G333.125$-$0.56) are shown in Figure \ref{fig:region_spectra}, showing that the signal-to-noise level is comparable among the molecular lines. From the spectral profiles in Figure \ref{fig:avespec} we can see that the molecular emission peaks at $\sim\,-50$ km s$^{-1}$, the systemic velocity of the G333 cloud, and spans a velocity range of $\sim\,-70$ to $\sim\,-40$ km s$^{-1}$. $^{13}$CO, C$^{18}$O and CS have similar velocity structure, such as the emission `shoulders' at $\sim\,-40$ and $\sim\,-55$ km s$^{-1}$. The $\sim\,-55$ km s$^{-1}$ emission feature also appears in the HCO$^+$ and HNC profiles. HCN and N$_2$H$^+$ have the widest velocity range, but this is due to hyperfine splitting rather than actual differences in distribution. The $^{13}$CO velocity feature centred at $-70$ km s$^{-1}$ comes from a different cloud along line of sight, as discussed in BWC2006.\\
%
%%% VELOCITY gradient %%%
\indent From the $^{13}$CO data, BWC2006 noted a linear velocity gradient of $\sim$ 0.2 km s$^{-1}$ pc$^{-1}$ across the GMC (1 arcmin is 1 pc for this GMC at distance of 3.6 kpc), which is five times larger than the velocity gradient due to Galactic rotation ($\sim0.04$ km s$^{-1}$ pc$^{-1}$ for an angular separation of $\sim$65 arcmin at $b = 333^{\circ}$, see BWC2006). To investigate whether molecules other than $^{13}$CO are involved in this bulk gas motion, we plotted the position-velocity (pv) map of CS, HCO$^+$, HNC and C$_2$H across the GMC as shown in Figure \ref{fig:pv_map}. The pv slice is positioned such that it cuts through the major emission ridge of the GMC, as shown in the top panel inset. The centre position (zero angular offset) is roughly at $\alpha_{J2000} = 16^h21^m$, $\delta_{J2000} = -50^d$30$^m$ near IRAS16172$-$5028, with positive offset corresponding to the southern part of the GMC. \\
\indent Evident in the CS pv map are a number of emission clumps; those near the GMC centre (centred on 0' offset) are surrounded by diffuse gas. The line widths of these emission clumps are $\sim$ 4 to 14 km s$^{-1}$. Similar to $^{13}$CO, the velocity gradient is also noticeable in CS, HCO$^+$, HNC and C$_2$H, $\sim$ 10 km s$^{-1}$ over roughly 50 arcmin ($\sim$ 0.2 km s$^{-1}$ pc$^{-1}$), comparable to that of the $^{13}$CO. The HCO$^+$ pv map shows similar overall structure to CS in general, however the centre velocity of the strong emission clump at $-30$ arcmin (G333.6$-$0.2, as labelled in Figure \ref{fig:pv_map}) is $-50$ km s$^{-1}$ for HCO$^+$, compared to $-48$ km s$^{-1}$ for CS, with HCO$^+$ emission being more extended towards the negative velocities ($\sim-60$ km s$^{-1}$) than CS. Where the CS emission clump peaks at $-48$ km s$^{-1}$, the HCO$^+$ emission clump is less intense and appears to host two components. In the HNC emission, the structure of this feature is consistent with CS. In the 3-D data cube, the HCO$^+$ spectrum shows a deep self-absorption feature at this position, when compared with the optically thin H$^{13}$CO$^+$, explaining the absence of HCO$^+$ emission. In general, the HNC and C$_2$H pv maps show a similar velocity structure to CS, except there is no detectable C$_2$H at $\sim$ 28 arcmin at $-58$ km s$^{-1}$.\\ 
\indent In the next two sections, we will present two different approaches in characterising the distribution of the emission in the GMC, clump finding with {\sc gaussclumps} and principal component analysis.\\
%
% ########################################################
%  clump analysis
% ########################################################
\section{Clump analysis}\label{sec:clumps}
One of the aims of this multi-molecular line mapping is to examine how different molecules correlate with each other in the GMC. To characterise the distribution of molecules, we have used the Gaussian clump decomposition algorithm {\sc gaussclumps} \citep{{Stutzki1990},{Kramer1998}} to decompose the observed three-dimensional data cubes into individual `clumps' of emission. There has been some discussion in recent years about the best type of automatic clump decomposition algorithm to use in giant molecular clouds (see \citeauthor{Sheth2008} \citeyear{Sheth2008} for a good discussion on this subject). {\sc gaussclumps} and {\sc clumpfind} \citep*{Williams1994} are algorithms which work by identifying peak pixels within a data cube or image, and then identifying other nearby pixels which may be part of the same clump.  {\sc clumpfind} stops adding pixels to a clump when it reaches pixels below a user specified contour level (see BWC2006 for a more lengthy discussion of this algorithm). {\sc gaussclumps} decomposes the data cube by iteratively fitting the peak position as a Gaussian distribution, then subtracting the fitted clumps and fitting the residual map. Both these methods have the disadvantage that they tend to break the emission into many small clumps close to the resolution element of the data set, particularly with a low signal-to-noise data set. \cite*{Rosolowsky2006} have described a different method, implemented by the algorithm {\sc cprops}, which uses a combination of moments and principal component analysis to identify clumps in data cubes, which they find is more robust against the effects of resolution and noise. However, of the three methods, we find {\sc gaussclumps} provided the best set of physically plausible clumps for the molecular transitions presented in this paper, which trace dense confined gas, with {\sc cprops} finding very few clumps in each data cube. This is consistent with the finding of \cite*{Rosolowsky2006} that {\sc gaussclumps} is the most effective algorithm for separating tight blends of clouds. Consequently, we have chosen to use the {\sc gaussclumps} algorithm, but have also undertaken a careful check of the clumps produced to confirm that they are physically plausible.\\
\indent The molecules selected for such clump finding are CS, HCO$^+$, HNC and C$_2$H. The selection is based on CS and HCO$^+$ emission not having hyperfine splitting, HNC hyperfine splitting not being resolvable, and the C$_2$H hyperfine component separation being so wide that only one component falls into the velocity range. The data cubes were smoothed with a hanning window width of five channels and binning window of two channels, giving a velocity resolution of $\sim0.3$ km s$^{-1}$, to improve the signal-to-noise ratio. The rms  levels after binning are 0.070, 0.065, 0.06 and 0.06 K (in antenna temperature, T$_A^*$) per 0.3 km s$^{-1}$ for CS, HCO$^+$, HNC and C$_2$H respectively. We have set the intensity threshold at the 5$\sigma$ level, therefore clumps with peak temperature below 5$\sigma$ are discarded for CS, HCO$^+$ and HNC. For C$_2$H, due to its comparatively weak emission, we only discarded clumps with peak temperature below the 3$\sigma$ level. For all molecules the spectral line emission towards the clumps was examined to confirm that the emission is real. We also excluded clumps that have angular sizes smaller than 1.5 beam-widths to reduce the number of false detections. Clumps that are found on the edges of the maps are also discarded, due to the lower sensitivity as the map edges contain one pass only. \\

\subsection{Clump properties}
\begin{figure*}
  \centering
  \includegraphics[]{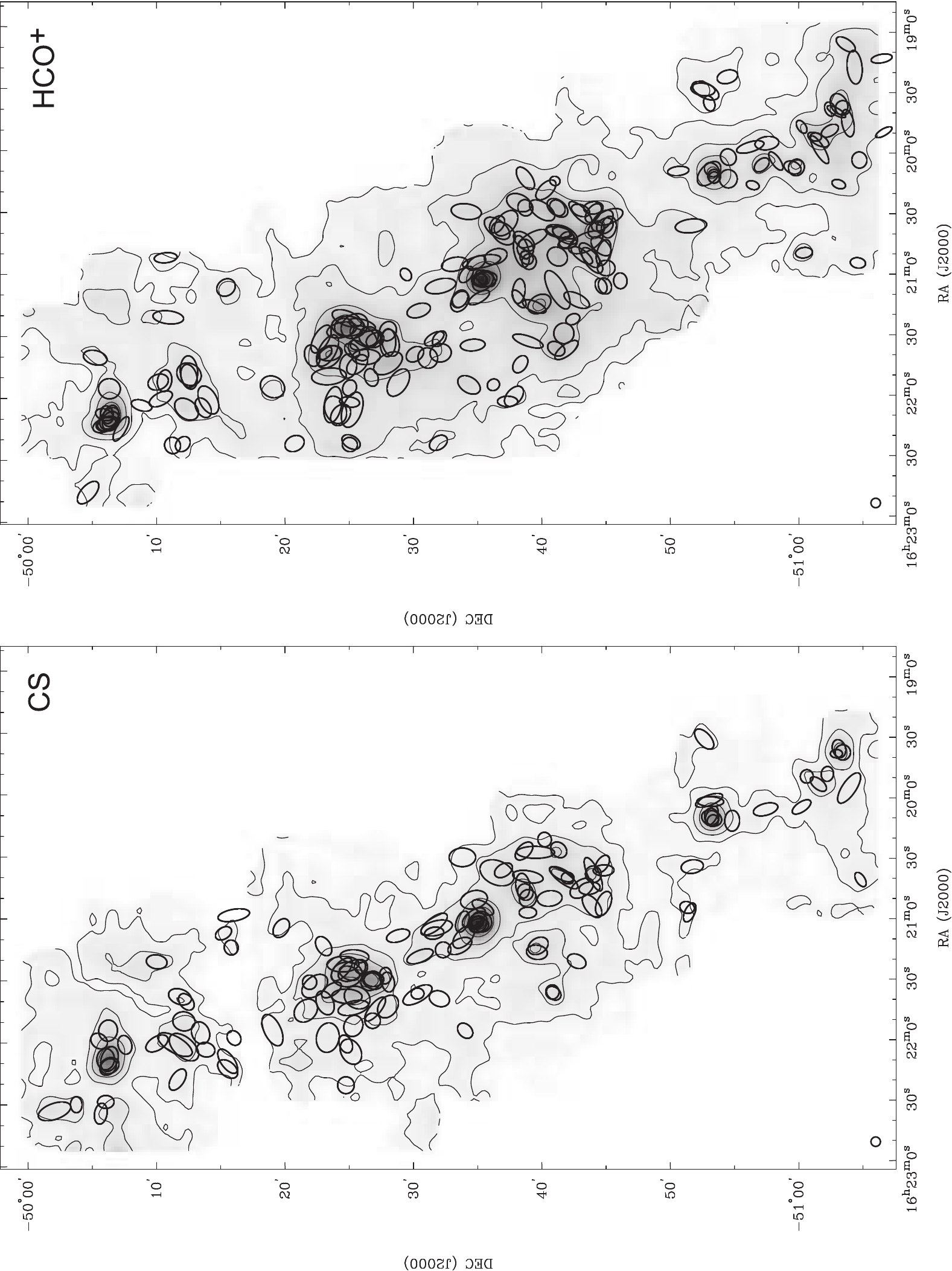}
  \caption[]{The integrated emission maps (contours and grey scale) of CS and HCO$^+$ (for the contour levels, see Figure
  \ref{fig:int1}) overlaid with the corresponding {\sc gaussclumps} clump fits of the 3-D data 
  cube. The ellipses show the orientation and size of the
  clumps. Refer to Tables \ref{tab:CS_gcl} and \ref{tab:HCOp_gcl} for details of the clumps identified.}
  \label{fig:gcl_map1}
\end{figure*}

\begin{figure*}
  \centering
  \includegraphics[]{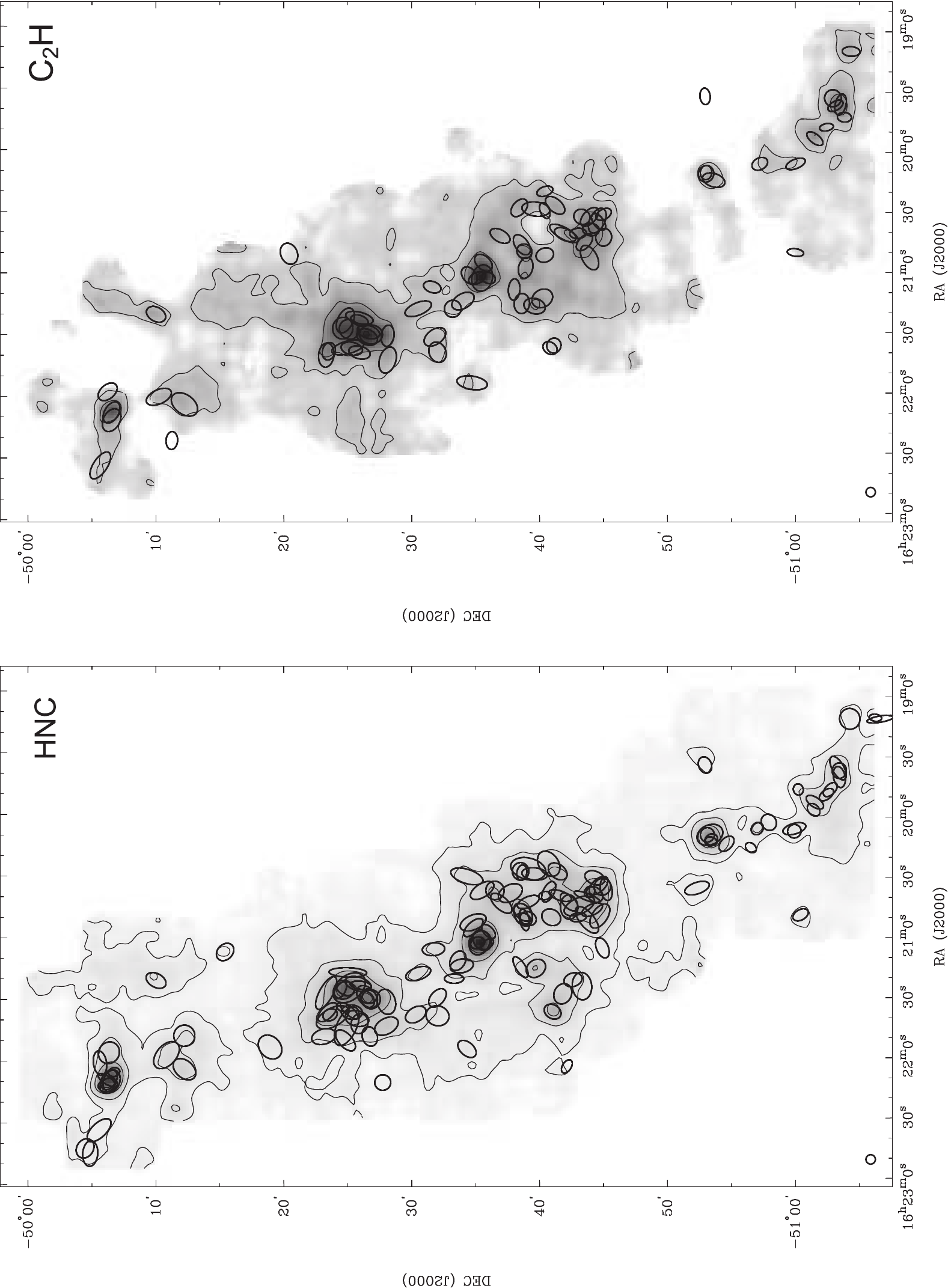}
  \caption[]{The integrated emission maps (contours and grey scale) of HNC and C$_2$H (for the contour levels, see Figure \ref{fig:int2}) overlaid with the corresponding {\sc gaussclumps}
  clump fits of the 3-D data cube. The ellipses show the orientation and size of the
  clumps. Refer to Tables \ref{tab:HNC_gcl}
  and \ref{tab:C2H_gcl} for details of the clumps identified.}
  \label{fig:gcl_map2}
\end{figure*}

{\sc gaussclumps} finds 129, 186, 128 and 78 clumps in the CS, HCO$^+$, HNC and C$_2$H data cubes, respectively. {\sc gaussclumps} reports centre positions and velocities, line widths, angular sizes, peak temperatures and internal velocity gradients of the identified clumps. We have listed the parameters of the brightest 20 clumps of CS, HCO$^+$, HNC and C$_2$H emission according to their peak intensities, along with the derived luminosity in Table \ref{tab:CS_gcl}, \ref{tab:HCOp_gcl}, \ref{tab:HNC_gcl} and \ref{tab:C2H_gcl}, respectively. A complete list of the clumps is available in the online version. The temperatures listed are corrected for the beam efficiency, T$_{b}$ = T$_A^*/\eta_{\nu}$, where $\eta_{\nu}$ is the extended beam efficiency. For the twenty brightest clumps of each molecular species, we also listed their associated clumps in other molecules in the last column of the tables. The association criteria are: the clumps coincide within one beam and the line widths overlap. We have also plotted the clump positions on the corresponding integrated intensity map in Figures \ref{fig:gcl_map1} and \ref{fig:gcl_map2}, with ellipses of size proportional to the angular sizes of the clumps. A summary of the clump properties is listed in Table \ref{tab:cl_stat} for comparison. The centroid velocities and line widths of the clumps among the four molecules are similar. The angular sizes of the clumps are also comparable among the four molecules, but note that the lower limit of angular size is just above the threshold value (1.5 beam-width) we have set. \\
%
%%% CS clumps %%%
\begin{table*}
  \caption[]{The 20 brightest CS clumps selected according to decreasing peak temperature with line intensities above the 
    5$\sigma$ level of the Hanning smoothed data cube. Also in Tables \ref{tab:HCOp_gcl},
    \ref{tab:HNC_gcl} and \ref{tab:C2H_gcl}: The listed properties
    are centre position (RA and Dec), centre velocity ($v$ in km s$^{-1}$), full width to
    half maximum (FWHM) line width ($\Delta V$ in km s$^{-1}$),
    FWHM of the two principal axes ($D_{\mathrm{x}}$ and
    $D_{\mathrm{y}}$ in arcmin), peak temperature ($T_{\mathrm{b}}=T^*_{\rm A} / \eta_{\nu}$ in
    K), internal velocity gradient across the clump ($dv/dr$ in km s$^{-1}$ arcmin$^{-1}$),
    calculated luminosity ($L$ in K km s$^{-1}$ pc$^2$),
    $L=(d\, \mathrm{[pc]})^2(\frac{\pi}{180 \times 3600})^2
    (r_x r_y) \int T_b \delta v$ (see text for details) 
    and associated clumps in other molecules.
    See the online version for a complete list of 129 clumps.}
  \label{tab:CS_gcl}
  \begin{tabular}{c c c c c c c c c c p{50mm}}
    \hline
    \# & RA & Dec & $v$ & $\Delta V$ & $D_{\mathrm{x}}$ & $D_{\mathrm{y}}$
     & Peak $T_{\mathrm{b}}$ & $dv/dr$ & $L$ & Associated clumps in other molecules \\
    \hline
    \hline \\[0.2ex]
    CS01 & 16:21:02.0 & -50:35:04.9 & -52.2 & 2.8 & 1.7 & 1.4 & 7.6 & 0.4 & 48.7
     & HCOp02, HNC02, C2H01 \\
    CS02 & 16:21:01.3 & -50:35:22.9 & -49.5 & 2.5 & 1.6 & 1.6 & 5.2 & 0.1 & 33.4
     & HCOp02, HNC02, HNC05, C2H01 \\
    CS03 & 16:22:07.8 & -50:06:22.9 & -47.8 & 4.1 & 2.0 & 1.3 & 4.9 & 0.4 & 49.1
     & CS14, HCOp03, HCOp09, HNC01, HNC10 \\
    CS04 & 16:21:02.0 & -50:34:58.9 & -55.3 & 3.1 & 1.0 & 1.2 & 4.4 & 0.1 & 14.8
     & HCOp05, C2H05 \\
    CS05 & 16:20:10.2 & -50:53:16.9 & -57.4 & 3.4 & 1.5 & 1.6 & 3.8 & 0.8 & 29.8
     & CS09, CS15, HCOp18, HNC11, HNC17, C2H06\\
    CS06 & 16:21:27.1 & -50:24:52.9 & -50.4 & 2.6 & 1.9 & 2.4 & 3.6 & 0.2 & 42.2
     & HCOp03, HNC03, C2H08 \\
    CS07 & 16:21:30.3 & -50:26:46.9 & -52.2 & 3.2 & 1.4 & 1.7 & 3.6 & 0.2 & 26.7
     & CS16, HCOp16, HCOp20, HNC12, C2H02 \\
    CS08 & 16:21:13.3 & -50:39:46.9 & -55.6 & 2.0 & 1.3 & 1.7 & 2.9 & 0.5 & 12.4
     & CS17, HCOp07, HNC07, C2H07 \\
    CS09 & 16:20:11.5 & -50:53:22.9 & -54.2 & 3.2 & 1.1 & 0.9 & 2.6 & 0.5 & 8.1
     & CS05, HCOp18, HNC17 \\
    CS10 & 16:20:26.0 & -50:41:22.9 & -56.0 & 2.4 & 1.2 & 1.4 & 2.6 & 0.3 & 9.9
     & HCOp14, HNC16, C2H11 \\
    CS11 & 16:20:46.9 & -50:38:46.9 & -53.3 & 2.6 & 1.4 & 1.2 & 2.5 & 0.3 & 10.4
     & HNC18, C2H04 \\
    CS12 & 16:21:37.8 & -50:24:52.9 & -50.8 & 3.0 & 1.1 & 2.4 & 2.0 & 0.5 & 16.1
     & C2H15 \\
    CS13 & 16:19:37.1 & -51:03:28.9 & -50.6 & 3.7 & 1.7 & 1.2 & 2.0 & 0.7 & 14.6
     & C2H12 \\
    CS14 & 16:22:07.8 & -50:06:04.9 & -45.3 & 2.0 & 1.7 & 1.4 & 2.0 & 0.3 & 8.5
     & CS03, HCOp04, HNC01, C2H03 \\
    CS15 & 16:20:09.6 & -50:53:16.9 & -60.1 & 2.4 & 1.4 & 1.4 & 2.0 & 0.6 & 8.6
     & CS05, HNC11 \\
    CS16 & 16:21:30.3 & -50:26:52.9 & -54.8 & 2.5 & 1.2 & 1.6 & 1.9 & 0.6 & 9.6
     & CS07, HCOp16, HCOp20, HNC12, C2H02 \\
    CS17 & 16:21:16.5 & -50:39:40.9 & -57.6 & 2.0 & 1.2 & 1.6 & 1.9 & 0.1 & 7.3
     & CS08, HCOp07, HNC07, C2H07 \\
    CS18 & 16:20:38.0 & -50:41:34.9 & -54.8 & 2.1 & 0.9 & 2.0 & 1.9 & 0.3 & 6.9 & $-$ \\
    CS19 & 16:21:18.3 & -50:30:34.9 & -52.2 & 2.0 & 0.9 & 2.2 & 1.9 & 0.4 & 7.1
     & C2H13 \\
    CS20 & 16:21:13.3 & -50:33:40.9 & -51.5 & 2.1 & 1.1 & 2.0 & 1.8 & 0.4 & 8.3
     & HNC15 \\
    \hline
  \end{tabular}
\end{table*}

%%% HCOp clumps %%%
%\setcounter{table}{2}
\begin{table*}
  \centering
  \caption[]{Same as Table \ref{tab:CS_gcl} but for the 20 brightest HCO$^+$ clumps with
    peak temperatures above 5$\sigma$ level of the Hanning smoothed data cube. The clumps
    are listed in descending peak T$_{\mathrm{b}}$ order. See the online version for a
    complete list of 186 clumps.}
  \label{tab:HCOp_gcl}
  \begin{tabular}{c c c c c c c c c c p{48mm}}
    \hline
    \# & RA & Dec & $v$ & $\Delta V$ & $D_{\mathrm{x}}$ & $D_{\mathrm{y}}$
     & Peak $T_{\mathrm{b}}$ & $dv/dr$ & $L$ & Associated clumps in other molecules \\
    \hline
    \hline \\[0.2ex]
    HCOp01 & 16:22:08.5 & -50:06:28.9 & -49.1 & 2.4 & 1.7 & 1.1 & 5.4 & 0.6 & 23.8
     & CS03, HNC01, HNC10, C2H03 \\
    HCOp02 & 16:21:03.2 & -50:35:28.9 & -51.4 & 3.8 & 1.4 & 1.8 & 4.9 & 0.6 & 47.3
     & CS01, CS02, HNC02, C2H01 \\
    HCOp03 & 16:21:25.8 & -50:24:40.9 & -50.1 & 2.8 & 1.4 & 1.5 & 4.4 & 0.5 & 26.1
     & CS06, HNC03, C2H08 \\
    HCOp04 & 16:22:09.7 & -50:06:22.9 & -46.7 & 1.7 & 1.9 & 1.2 & 3.0 & 0.2 & 11.6
     & CS03, CS14, HNC01, C2H03 \\
    HCOp05 & 16:21:02.6 & -50:35:10.9 & -55.5 & 3.2 & 0.9 & 1.2 & 2.9 & 0.1 & 9.8
     & CS04, C2H05 \\
    HCOp06 & 16:20:05.8 & -50:57:10.9 & -56.0 & 2.0 & 1.1 & 1.4 & 2.9 & 0.6 & 8.5
     & HNC13, C2H20 \\
    HCOp07 & 16:21:14.6 & -50:39:52.9 & -56.0 & 3.1 & 1.5 & 1.9 & 2.9 & 0.8 & 23.6
     & CS08, CS17, HNC07, C2H07 \\
    HCOp08 & 16:21:40.3 & -50:23:28.9 & -50.8 & 1.9 & 1.4 & 2.1 & 2.6 & 0.3 & 13.8
     & HNC08 \\
    HCOp09 & 16:22:11.0 & -50:06:28.9 & -51.5 & 2.8 & 1.6 & 1.2 & 2.5 & 0.3 & 13.6
     & CS03, HCOp01, HNC10 \\
    HCOp10 & 16:20:38.1 & -50:44:16.9 & -53.1 & 2.4 & 1.2 & 1.4 & 2.3 & 0.4 & 8.4
     & HNC04, C2H10 \\
    HCOp11 & 16:20:44.4 & -50:43:28.9 & -60.5 & 1.7 & 1.6 & 2.9 & 2.1 & 0.2 & 15.1
     & HNC14 \\
    HCOp12 & 16:21:36.7 & -50:41:10.9 & -59.8 & 2.5 & 1.1 & 1.4 & 2.1 & 0.5 & 8.0
     & HNC09 \\
    HCOp13 & 16:20:39.3 & -50:41:34.9 & -54.6 & 2.0 & 1.0 & 1.4 & 2.0 & 0.6 & 5.1
     & CS18, C2H16 \\
    HCOp14 & 16:20:27.3 & -50:41:16.9 & -55.3 & 2.3 & 1.3 & 1.8 & 1.9 & 0.5 & 10.1
     & CS10, HNC16, C2H11 \\
    HCOp15 & 16:21:43.5 & -50:28:04.9 & -50.1 & 2.0 & 1.5 & 2.5 & 1.8 & 0.2 & 12.8
     & HNC20, C2H17 \\
    HCOp16 & 16:21:32.2 & -50:27:04.9 & -52.4 & 2.4 & 1.1 & 1.4 & 1.8 & 0.3 & 6.5
     & CS07, CS16, HNC12, C2H02 \\
    HCOp17 & 16:20:53.1 & -50:43:58.9 & -54.0 & 2.2 & 1.2 & 2.4 & 1.8 & 0.6 & 10.6
     & HNC19 \\
    HCOp18 & 16:20:11.5 & -50:53:28.9 & -54.4 & 2.6 & 1.2 & 1.1 & 1.8 & 0.4 & 5.7
     & CS05, CS09, HNC17 \\
    HCOp19 & 16:19:48.6 & -51:02:28.9 & -58.9 & 2.0 & 1.2 & 1.5 & 1.8 & 0.1 & 5.6
     & $-$ \\
    HCOp20 & 16:21:32.1 & -50:26:22.9 & -49.7 & 1.9 & 1.2 & 2.6 & 1.7 & 0.3 & 10.3
     & CS07, CS16, HNC12 \\
    \hline
  \end{tabular}
\end{table*}

%%% HNC clumps %%%
%\setcounter{table}{3}
\begin{table*}
  \centering
  \caption[]{Same as Table \ref{tab:CS_gcl} but for the 20 brightest HNC clumps with peak
    temperature above 5$\sigma$ level of the Hanning smoothed data cube. The clumps are
    listed in descending T$_{\mathrm{b}}$ order. See the online version for a complete
    list of 128 clumps.}
  \label{tab:HNC_gcl}
  \begin{tabular}{c c c c c c c c c c p{48mm}}
    \hline
    \# & RA & Dec & $v$ & $\Delta V$ & $D_{\mathrm{x}}$ & $D_{\mathrm{y}}$
     & Peak $T_{\mathrm{b}}$ & $dv/dr$ & $L$ & Associated clumps in other molecules \\
    \hline
    \hline \\[0.2ex]
    HNC01 & 16:22:09.1 & -50:06:22.9 & -47.1 & 3.7 & 1.1 & 1.7 & 4.4 & 0.2 & 28.1
     & CS03, CS14, HCOp01, HCOp04, C2H03 \\
    HNC02 & 16:21:02.6 & -50:35:22.9 & -51.7 & 3.0 & 1.6 & 2.0 & 4.2 & 0.4 & 35.4
     & CS01, CS02, HCOp02, HNC05, C2H01 \\
    HNC03 & 16:21:25.8 & -50:24:46.9 & -50.4 & 2.9 & 1.3 & 1.6 & 3.4 & 0.5 & 19.8
     & CS06, HCOp03, C2H08 \\
    HNC04 & 16:20:38.7 & -50:44:22.9 & -52.7 & 2.4 & 1.2 & 1.4 & 2.9 & 0.5 & 10.9
     & HCOp10, C2H10 \\
    HNC05 & 16:21:01.3 & -50:35:34.9 & -49.2 & 2.1 & 1.4 & 1.7 & 2.8 & 0.2 & 13.2
     & CS02, HCOp02, HNC02, C2H01, C2H18 \\
    HNC06 & 16:21:02.6 & -50:35:10.9 & -54.6 & 2.6 & 0.9 & 1.2 & 2.8 & 0.1 & 7.4
     & CS04, HCOp02, HCOp05, HNC02, C2H05 \\
    HNC07 & 16:21:15.2 & -50:39:46.9 & -56.1 & 3.1 & 1.4 & 1.6 & 2.4 & 1.1 & 15.2
     & CS08, CS17, HCOp07, C2H07 \\
    HNC08 & 16:21:39.1 & -50:23:52.9 & -50.8 & 2.3 & 1.4 & 2.8 & 2.3 & 0.1 & 19.1
     & CS12, HCOp08, C2H15 \\
    HNC09 & 16:21:36.7 & -50:41:04.9 & -59.2 & 3.1 & 1.2 & 1.4 & 2.2 & 1.1 & 10.6
     & HCOp12 \\
    HNC10 & 16:22:09.7 & -50:06:34.9 & -49.7 & 2.8 & 1.8 & 1.3 & 2.2 & 0.3 & 12.9
     & CS03, HCOp01, HCOp09, C2H03 \\
    HNC11 & 16:20:08.9 & -50:53:22.9 & -58.1 & 3.3 & 1.9 & 2.1 & 2.1 & 0.6 & 25.7
     & CS05, CS15, C2H06 \\
    HNC12 & 16:21:30.3 & -50:26:52.9 & -52.1 & 3.0 & 1.5 & 1.5 & 2.1 & 0.3 & 12.6
     & CS07, CS16, HCOp16, HCOp20, C2H02 \\
    HNC13 & 16:20:05.8 & -50:57:04.9 & -56.1 & 2.0 & 1.1 & 1.0 & 2.0 & 0.6 & 4.3
     & HCOp06, C2H20 \\
    HNC14 & 16:20:44.4 & -50:43:10.9 & -60.5 & 1.7 & 1.4 & 2.5 & 1.9 & 0.1 & 10.7
     & HCOp11 \\
    HNC15 & 16:21:13.9 & -50:33:58.9 & -51.5 & 2.0 & 1.1 & 1.8 & 1.8 & 0.4 & 7.1
     & CS20 \\
    HNC16 & 16:20:27.3 & -50:41:28.9 & -56.1 & 2.3 & 1.2 & 1.7 & 1.8 & 0.2 & 7.9
     & CS10, HCOp14, C2H11 \\
    HNC17 & 16:20:11.5 & -50:53:28.9 & -54.6 & 2.9 & 1.1 & 1.1 & 1.8 & 0.5 & 5.6
     & CS05, CS09, HCOp18 \\
    HNC18 & 16:20:47.5 & -50:38:46.9 & -53.1 & 2.0 & 1.2 & 1.7 & 1.8 & 0.0 & 6.8
     & CS11, C2H04 \\
    HNC19 & 16:20:51.7 & -50:43:46.9 & -54.2 & 2.0 & 1.3 & 2.5 & 1.7 & 0.4 & 10.0
     & HCOp17 \\
    HNC20 & 16:21:43.5 & -50:28:04.9 & -50.3 & 1.9 & 1.2 & 2.0 & 1.6 & 0.1 & 6.9
     & HCOp15, C2H17 \\
    \hline
  \end{tabular}
\end{table*}

%%% C2H clumps %%%
%\setcounter{table}{4}
\begin{table*}
  \centering
  \caption[]{Same as Table \ref{tab:CS_gcl} but for the 20 brightest C$_2$H clumps with
    peak temperature above 3$\sigma$ level of the Hanning smoothed data cube. The clumps
    are listed in descending peak T$_{\mathrm{b}}$ order. See the online version for a
    complete list of 78 clumps. The C$_2$H clumps presented here refer to the
    main hyperfine component only, hence the luminosity is not calculated.}
  \label{tab:C2H_gcl}
  \begin{tabular}{c c c c c c c c c p{50mm}}
    \hline
    \# & RA & Dec & $v$ & $\Delta V$ & $D_{\mathrm{x}}$ & $D_{\mathrm{y}}$
     & Peak $T_{\mathrm{b}}$ & $dv/dr$ & Associated clumps in other molecules \\
    \hline
    \hline \\[0.2ex]
    C2H01 & 16:21:02.6 & -50:35:34.9 & -50.8 & 2.8 & 2.2 & 1.6 & 1.8 & 0.5
     & CS01, CS02, HCOp02, HNC02, HNC05, C2H18 \\
    C2H02 & 16:21:30.3 & -50:26:52.9 & -52.6 & 2.4 & 1.5 & 1.8 & 1.6 & 0.2
     & CS07, CS16, HCOp16, HNC12 \\
    C2H03 & 16:22:07.8 & -50:06:34.9 & -47.8 & 3.2 & 1.1 & 1.8 & 1.6 & 0.2
     & CS03, CS14, HCOp01, HCOp04, HNC01, HNC10 \\
    C2H04 & 16:20:49.4 & -50:38:52.9 & -53.7 & 2.1 & 1.2 & 1.2 & 1.4 & 0.3
     & CS11, HNC18 \\
    C2H05 & 16:21:02.6 & -50:35:10.9 & -54.0 & 2.1 & 0.9 & 1.4 & 1.3 & 0.3
     & CS04, HCOp02, HCOp05, HNC02, HNC06 \\
    C2H06 & 16:20:10.8 & -50:53:04.9 & -58.3 & 2.7 & 1.6 & 1.2 & 1.1 & 0.2
     & CS05, HNC11 \\
    C2H07 & 16:21:16.5 & -50:39:40.9 & -56.7 & 2.3 & 1.3 & 1.7 & 1.1 & 0.9
     & CS08, CS17, HCOp07, HNC07 \\
    C2H08 & 16:21:27.1 & -50:24:40.9 & -50.4 & 1.8 & 1.6 & 1.7 & 1.1 & 0.5
     & CS06, HCOp03, HNC03 \\
    C2H09 & 16:19:53.0 & -51:01:34.9 & -47.6 & 2.6 & 1.0 & 1.5 & 1.0 & 0.4
     & $-$ \\
    C2H10 & 16:20:38.7 & -50:44:04.9 & -52.6 & 2.0 & 1.1 & 1.3 & 1.0 & 0.7
     & HCOp10, HNC04 \\
    C2H11 & 16:20:26.7 & -50:41:16.9 & -55.5 & 2.4 & 1.1 & 1.8 & 1.0 & 0.5
     & CS10, HCOp14, HNC16 \\
    C2H12 & 16:19:35.8 & -51:03:34.9 & -50.1 & 2.5 & 1.2 & 1.7 & 1.0 & 0.2
     & CS13 \\
    C2H13 & 16:21:17.7 & -50:30:34.9 & -52.4 & 1.8 & 1.0 & 2.2 & 0.9 & 0.3
     & CS19 \\
    C2H14 & 16:22:04.1 & -50:12:16.9 & -49.7 & 2.0 & 1.5 & 2.3 & 0.9 & 0.4
     & $-$ \\
    C2H15 & 16:21:36.5 & -50:25:04.9 & -50.8 & 2.0 & 1.0 & 2.6 & 0.9 & 0.2
     & CS12, HNC08 \\
    C2H16 & 16:20:41.2 & -50:42:04.9 & -52.9 & 1.8 & 1.0 & 1.8 & 0.8 & 0.1
     & CS18, HCOp13 \\
    C2H17 & 16:21:42.9 & -50:28:10.9 & -50.5 & 1.4 & 1.3 & 2.1 & 0.8 & 0.3
     & HCOp15, HNC20 \\
    C2H18 & 16:21:02.6 & -50:36:04.9 & -48.8 & 1.3 & 0.9 & 1.1 & 0.8 & 0.3
     & CS05, CS09, HCOp02, HNC05 \\
    C2H19 & 16:21:20.0 & -50:10:04.9 & -43.5 & 1.9 & 1.1 & 1.6 & 0.8 & 0.1
     & $-$ \\
    C2H20 & 16:20:05.8 & -50:57:16.9 & -56.0 & 1.8 & 1.2 & 1.3 & 0.7 & 0.6
     & HCOp06, HNC13 \\
    \hline
  \end{tabular}
\end{table*}
\begin{table}
  \centering
  \caption{A summary of the clump properties for the CS, HCO$^+$, HNC and C$_2$H data cubes, derived from {\sc gaussclumps} (see Section \ref{sec:clumps}).}
  \label{tab:cl_stat}
  \begin{tabular}{c c c c c}%{|c| c c c c |}
    \hline% \\[0.2ex]
     & CS & HCO$^+$ & HNC & C$_2$H \\
    \hline \hline% \\[0.2ex]
    Number of clumps & 129 & 186 & 128 & 78 \\
    \hline
    \multicolumn{5}{c}{Centre velocity \scriptsize (km s$^{-1}$)}\\
    \hline
    lowest & $-61.9$ & $-64.8$ & $-65.9$ & $-61.4$ \\
    highest & $-44.0$ & $-44.4$ & $-44.4$ & $-43.5$ \\
    mean & $-52.0$ & $-52.6$ & $-52.6$ & $-52.4$ \\
    $\sigma$ & 4.1 & 3.8 & 3.7 & 3.5 \\
    \hline
    \multicolumn{5}{c}{$\Theta$ \scriptsize (arcmin)}\\
    \hline
    minimum & 0.9 & 0.8 & 0.8 & 0.9 \\
    maximum & 2.3 & 2.0 & 2.3 & 1.8 \\
    mean & 1.4 & 1.4 & 1.3 & 1.3 \\
    $\sigma$ & 0.2 & 0.2 & 0.3 & 0.2 \\
    \hline
    \multicolumn{5}{c}{Peak T$_{\mathrm{b}}$ \scriptsize (K)}\\
    \hline
    minimum & 0.7 & 0.7 & 0.7 & 0.5 \\
    maximum & 7.6 & 5.4 & 4.4 & 1.8 \\
    mean & 1.3 & 1.6 & 1.8 & 0.7 \\
    $\sigma$ & 1.0 & 0.7 & 0.7 & 0.3 \\
    \hline
    \multicolumn{5}{c}{dV \scriptsize (km s$^{-1}$)}\\
    \hline
    minimum & 0.4 & 0.5 & 0.4 & 0.3 \\
    maximum & 4.1 & 3.8 & 3.7 & 3.2 \\
    mean & 1.7 & 1.1 & 1.1 & 1.5 \\
    $\sigma$ & 0.7 & 0.6 & 0.6 & 0.6 \\
    \hline
    \multicolumn{5}{c}{Luminosity \scriptsize (K km s$^{-1}$ pc$^2$)} \\
    \hline
    minimum & 0.05 & 0.04 & 0.02 & 0.03 \\
    maximum & 5.2 & 5.0 & 4.0 & 1.8 \\
    mean & 0.6 & 0.4 & 0.5 & 0.2 \\
    $\sigma$ & 0.9 & 0.5 & 0.6 & 0.3 \\
    \hline
  \end{tabular}
\end{table}
\subsection{Optical depth and column density}\label{sec:opac}
\begin{figure}
  \centering
  \includegraphics[]{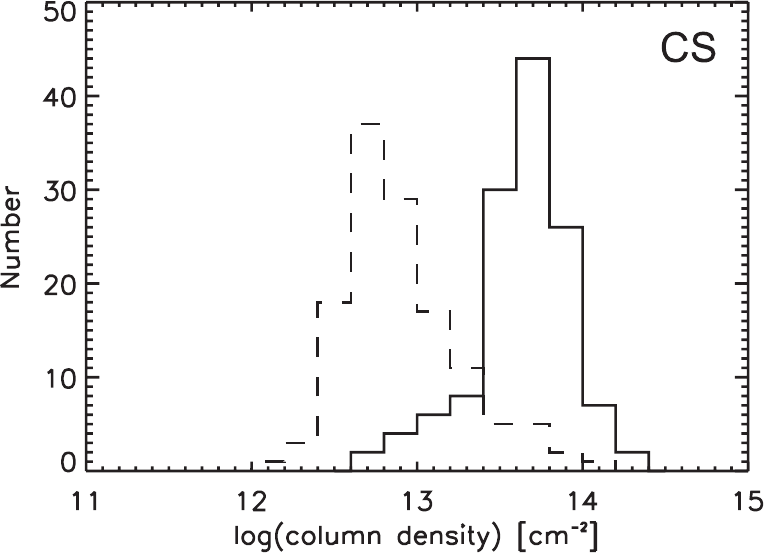}
  \caption[]{Histogram of CS clump column density with optical depth correction (solid).
  The dashed line shows the distribution obtained by assuming that CS is optically thin and has an
  excitation temperature of 20 K.}
  \label{fig:CS_Nopac}
\end{figure}

\begin{figure}
  \centering
  \includegraphics[]{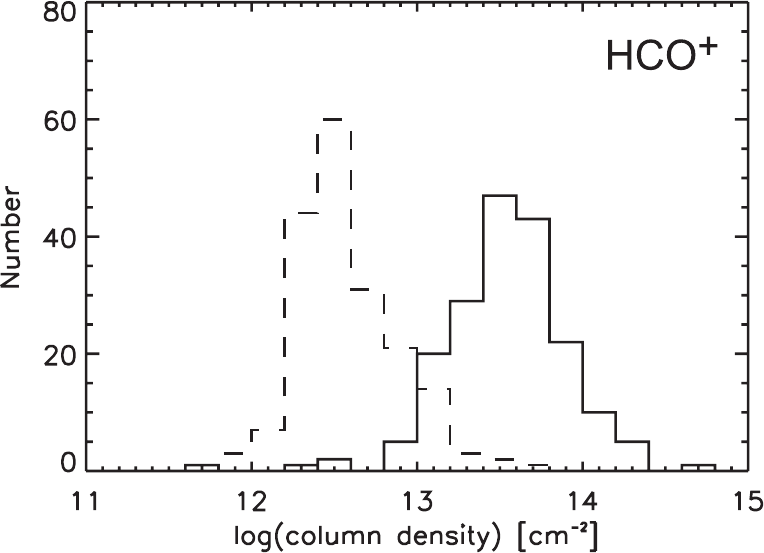}
  \caption[]{Histogram of HCO$^+$ clump column density with optical depth correction (solid).
  The dashed line shows the distribution obtained by assuming that HCO$^+$ is optically thin and has an
  excitation temperature of 20 K.}
  \label{fig:HCOp_Nopac}
\end{figure}

\begin{figure}
  \centering
  \includegraphics[]{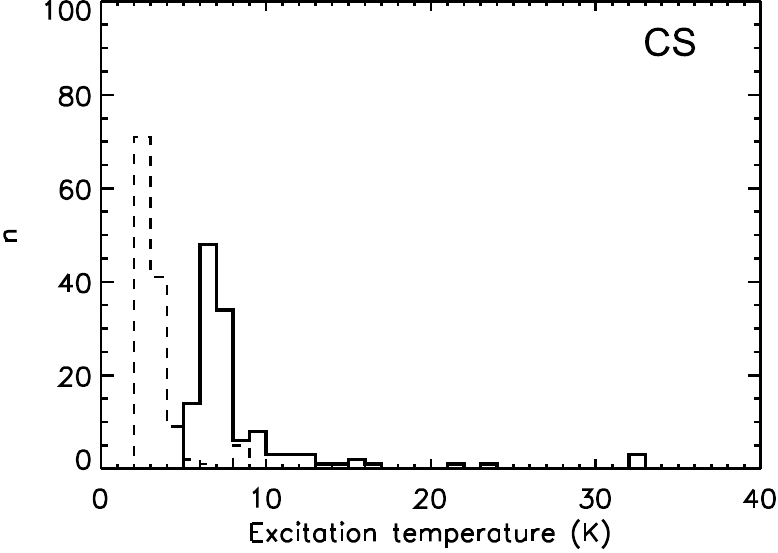}
  \caption[]{Histogram of CS clump excitation temperature with a beam filling factor of 0.2 (solid line). 
  The dashed line shows the distribution assuming a unity beam filling factor.}
  \label{fig:CS_Tex}
\end{figure}

\begin{figure}
  \centering
  \includegraphics[]{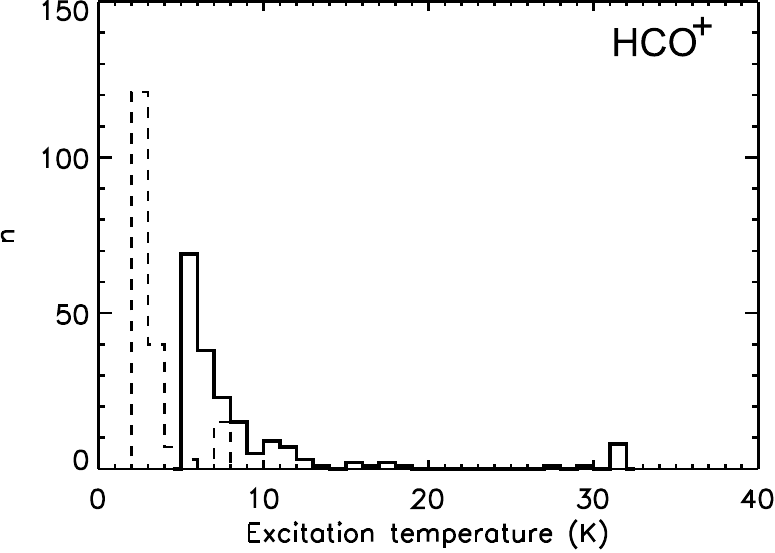}
  \caption[]{Histogram of HCO$^+$ clump excitation temperature with a beam filling factor of 0.2 (solid line). The dashed line shows the distribution assuming a unity beam filling factor.}
  \label{fig:HCOp_Tex}
\end{figure}

Among the four molecules decomposed by {\sc gaussclumps}, CS and HCO$^+$ have lines from rare isotopes, so-called `isotopologues' (C$^{34}$S and H$^{13}$CO$^+$) in our data set. Therefore it is possible to estimate optical depths for these molecules. Assuming identical excitation temperatures and optically thin emission from the rare isotopologues, the opacity $\tau$ relates to the brightness (beam efficiency corrected) of optically thick and thin emission lines as follows:
\begin{equation}\label{equ:tau}
    \frac{T_{\mathrm{b}}\mathrm{(thick)}}{T_{\mathrm{b}}\mathrm{(thin)}}
     =\frac{1-e^{-\tau_{\mathrm{thick}}}}{1-e^{-\tau_{\mathrm{thin}}}}
     = \frac{1-e^{-\tau_{\mathrm{thick}}}}{1-e^{-\tau_{\mathrm{thick}}/X}},
\end{equation}
where $X$ is the abundance ratio, assuming [CS/C$^{34}$S] = 22.5 \citep{Chin1996} and [HCO$^+$/H$^{13}$CO$^+$] = 50, based on [CO/$^{13}$CO] from \citet*{Wilson1994}. \\
\indent After solving for $\tau$, we next calculate the excitation temperature $T_{\mathrm{ex}}$ from the optically thick emission lines, CS and HCO$^+$. From the equation of radiative transfer, $T_{\mathrm{ex}}$ can be derived from
\begin{equation}\label{equ:Tex}
  T_{\mathrm{b}} = f[J(T_{\mathrm{ex}})-J(T_{\mathrm{bg}})][1-e^{-\tau}].
\end{equation}
$f$ is the beam filling factor which we assume to be 1, and $J(T) = T_0/(e^{T_0/T} - 1)$. Once $\tau$ and $T_{\mathrm{ex}}$ are known, the column density of the optically thick line can be derived with (cf \citeauthor{Purcell2006} \citeyear{Purcell2006})
\begin{equation}\label{equ:N}
  N = \frac{8 k \pi \nu^2}{h c^3 g_{\mathrm{u}} A_{\mathrm{ul}}}\, \frac{\tau}{1-e^{-\tau}}\,
    e^{-E_{\mathrm{u}}/k T_{\mathrm{ex}}}\, Q(T)\, \int T_{\mathrm{b}} dv ,
\end{equation}
where $A_{\mathrm{ul}}$ is the Einstein A coefficient in s$^{-1}$, $\nu$ is the transition frequency in Hz, $\int T_{\mathrm{b}} dv$ is the integrated brightness temperature in K km s$^{-1}$, $Q(T)$ is the partition function, $g_{\mathrm{u}}$ is the upper state degeneracy and $E_{\mathrm{u}}$ is the upper state energy in m$^{-1}$. At positions, where no C$^{34}$S or H$^{13}$CO$^+$ was detected, we have assumed also that the main isotopologue was optically thin and the excitation temperature was set to 3 K. Eleven CS and two HCO$^+$ clumps fall into this category. The choice of 3 K for the excitation temperature was motivated by the median value of $T_{\mathrm{ex}}$ from clumps that show emission in the main and the rare species (Figures \ref{fig:CS_Tex} and \ref{fig:HCOp_Tex}).\\
\indent Shown in Figure \ref{fig:CS_Nopac} and \ref{fig:HCOp_Nopac} are histogram plots of CS and HCO$^+$ clump column density. The solid line shows the column density after correction for optical depth, and the dashed line shows the derived column density distribution by assuming CS and HCO$^+$ are optically thin with an excitation temperature of 20 K. The plots show that after correction, the column density increases by an order of magnitude. An examination of the clump spectra confirms that clumps with column density below 10$^{13}$ cm$^{-2}$ are those with minimal or no optically thin line detections, while those with column density of 10$^{14}$ cm$^{-2}$ or above are optically thick clumps. \\
\indent An excitation temperature of 3 K is low. One possible cause could be due to the optically thin lines (C$^{34}$S and H$^{13}$CO$^+$) being spatially compact, and the assumption of having the same beam efficiency as the more common isotopologue may not hold. We  thus also calculated the excitation temperatures of both CS and HCO$^+$ with extended beam efficiency ($\eta_{\mathrm{xb}}$) for CS and HCO$^+$, and main beam efficiency ($\eta_{\mathrm{mb}}$) for C$^{34}$S and H$^{13}$CO$^+$. The excitation temperatures of CS and HCO$^+$ then increases to $\sim$4 K, still low. Another possibility is that the clumps are fragmented and unresolvable within the Mopra beam ($\sim$36 arcsec at 100-GHz). The clump finding algorithm sees these fragmented clumps as one single clump. Thus, the assumption of the beam filling factor ($f$) being unity does not hold, despite the fact that all identified clumps are larger than 1.5 beam widths. Various studies show that CS clumps are typically fragmented, with a volume filling factor as low as 0.2 \citep*[e.g.][]{{Stutzki1990},{Juvela1998}}. To investigate the effect of beam filling factor on the excitation temperature, we derived the excitation temperatures of CS and HCO$^+$ clumps with various beam filling factors. We found that with $f = 0.2$, approximately 74 per cent of the CS clumps, and 56 per cent of the HCO$^+$ clumps have $T_{\mathrm{ex}}$ between 6 to 20 K. Histograms of CS and HCO$^+$ excitation temperatures are shown in Figure \ref{fig:CS_Tex} and \ref{fig:HCOp_Tex} to illustrate the effect of lowering beam filling factor. The mean excitation temperature increases greatly by lowering the beam filling factor, suggesting the clumps may indeed fragmented and not resolvable with MOPRA beam. This would provide an explanation for the lack of correlation between clump size and line width (see Section \ref{sec:larson} for details).\\

%
% ########################################################
%  PCA
% ########################################################
\section{Principal component analysis}\label{sec:PCA}
\begin{figure}
  \centering
  \includegraphics[]{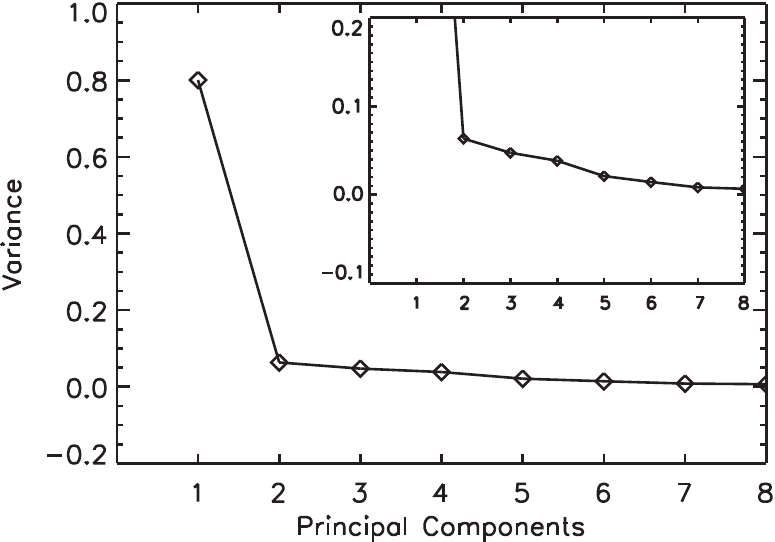}
  \caption[]{Plot of variances of the principal components. The inset presents a 
  zoomed version of the plot, showing the second and higher order components in more detail.}
  \label{fig:PCA_scree}
\end{figure}
\begin{figure}
  \centering
  \includegraphics[]{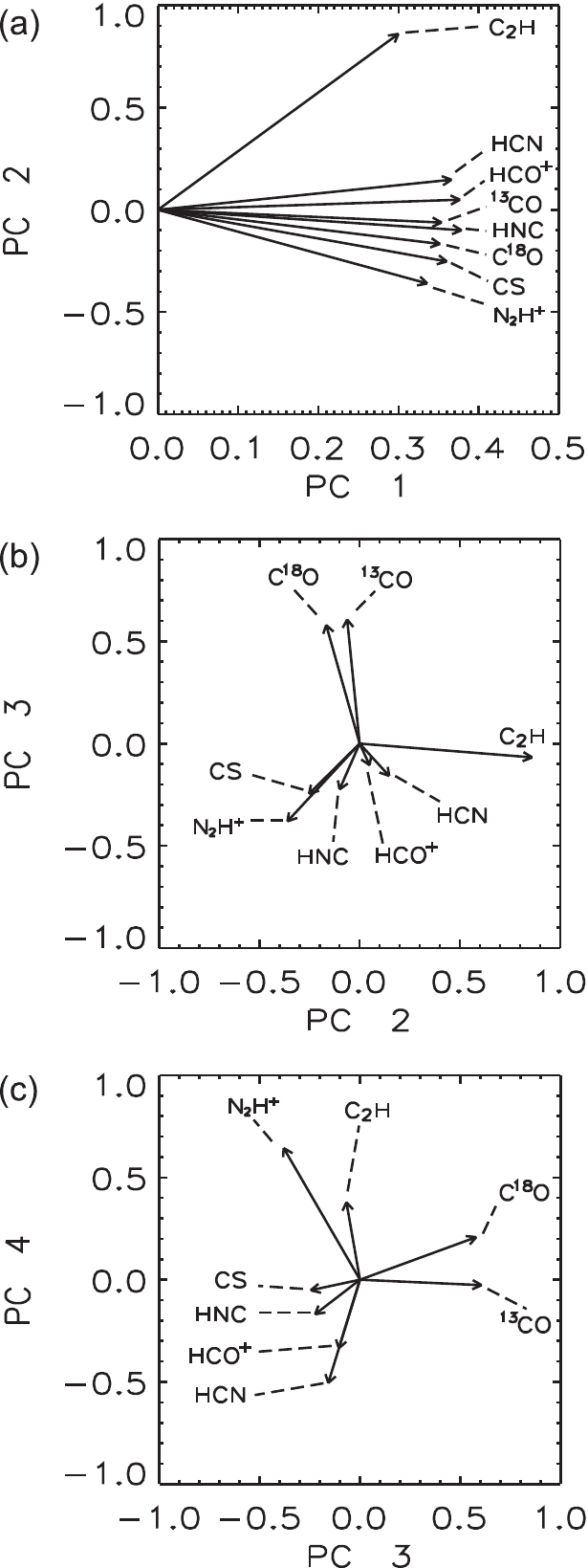}
  \caption[]{Plots of eigenvectors of the first three principal components of the data set. Each of
   the eigenvectors represent the component of that molecule in the relevant principal component.}
  \label{fig:evec}
\end{figure}
\begin{figure*}
  \centering
  \includegraphics[]{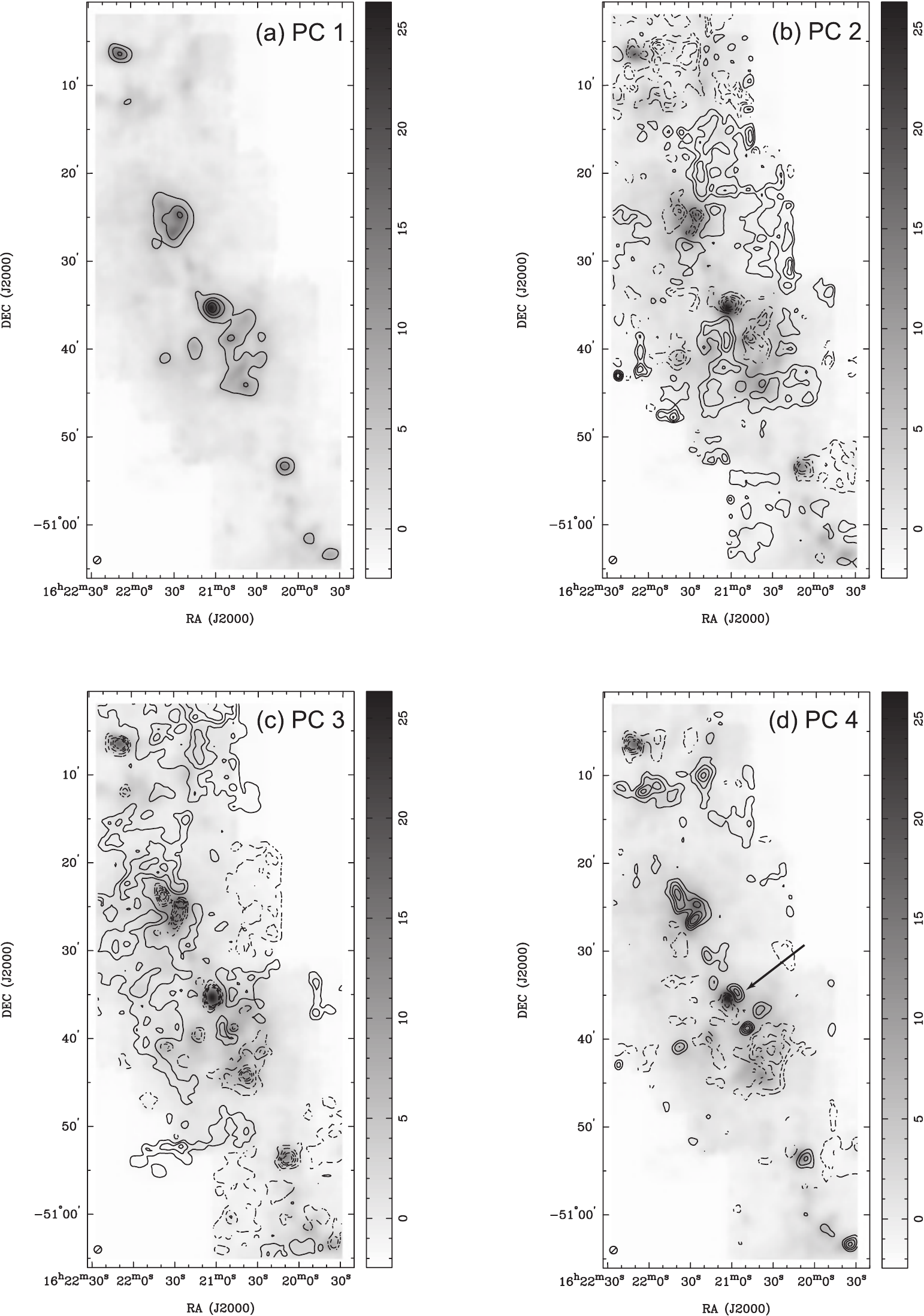}
  \caption[]{Constructed images of the first four principal components of the GMC. The
   contour levels are multiples of 20 per cent of the peak. The dashed contours represent
   an anti-correlation, while the solid contours represent a positive correlation in each
   of the principal components.
   The grey scale in each of the images shows the first principal component,
   overlaid with contours of the other principal components for ease of comparison. The
   lower left circles of each images indicate the beam size. The arrow indicates an example
   of anti-correlation between HCO$^+$ and N$_2$H$^+$ as discussed in Section
   \ref{sec:pca_N2Hp_HCOp}.}
  \label{fig:pc1_4}
\end{figure*}

An alternative way to study the distribution of the molecules is by performing principal component analysis (PCA) on the integrated emission maps of the region. PCA is a multivariate data analysis technique, its purpose being to reduce the dimensionality of a data set. Mathematically the derivation of principal components involves finding the eigenvalues and eigenvectors of a covariance or correlation matrix \citep[see e.g.][and references therein]{{Jolliffe2002},{Heyer1997},{Ungerechts1997}}. \\
\indent The purpose of performing PCA on the integrated emission maps is to characterise differences in molecular distribution. The G333 cloud consists of star forming sites at different stages of evolution, from cold dense starless cores to H{\sc ii} regions. Hence we expect to see chemical differences across the cloud. The data set covers varieties of chemical probes. Representative ones are the outflow tracer HCO$^+$, the large scale gas tracer $^{13}$CO, the dense gas tracer CS, and the cold dense gas tracer N$_2$H$^+$. Interesting aspects are chemical differences and a possible temperature dependance of HCN to HNC abundance ratio. In this section, we will discuss the result of a PCA decomposition on the whole G333 molecular cloud; the PCA of individual regions of interest will be discussed in later sections. \\
\indent We have chosen eight molecule lines with high signal-to-noise ratio for this analysis: $^{13}$CO, C$^{18}$O, CS, HCO$^+$, HCN, HNC, N$_2$H$^+$ and C$_2$H (Figures \ref{fig:int1} and \ref{fig:int2}). Among these, C$_2$H has the lowest signal-to-noise level and $^{13}$CO has the highest. The data sets were imported into {\sc idl} (a data visualisation and analysis platform) and each of their standard deviations were obtained. We decided to derive the principal components of the correlation matrix, rather than the covariance matrix. This is to avoid the principal components being dominated by a single variable \citep*[][Chapter 3.3]{Jolliffe2002}. An alternative approach involves normalising the data set first, and then deriving the principal components from the covariance matrix. We in fact derived the principal components using these two different approaches and obtained essentially the same result. \\
\indent Listed in Table \ref{tab:cor_matrix} is the correlation matrix of the input molecules, which describes how well the molecules correlate with each other. The correlation matrix shows there are four pairs of molecules which have correlation coefficients equal to or above 0.9. These are HCO$^+$ and HCN, HCO$^+$ and HNC, HCN and HNC, HNC and CS. This is consistent with the integrated intensity maps shown in Figure \ref{fig:int1} and \ref{fig:int2}, where CS, HCO$^+$, HCN and HNC have similar large scale spatial distribution. However CS does show small differences in distribution, as discussed in Section \ref{sec:maps}. Among the eight molecules, C$_2$H and N$_2$H$^+$ have the lowest correlation coefficient. After forming the correlation matrix of the data set, we derived the eigenvalues and eigenvectors of the correlation matrix (Table \ref{tab:evectors_val}). The eigenvalues indicate that the first principal component accounts for 80 per cent of total variation, while the first three principal components together account for over 90 per cent of total variation in the data set (see Table \ref{tab:evectors_val} and Figure \ref{fig:PCA_scree}). Figure \ref{fig:PCA_scree} also shows that the first four principal components contain features above the noise level, while the variations contained in higher components are insignificant.\\
\begin{table*}
  \centering
  \caption{The correlation matrix of the input molecular data set for principal component
    analysis.}
  \label{tab:cor_matrix}
  \begin{tabular}{c c c c c c c c c}
    \hline% \\[0.2ex]
     & CS & HCO$^+$ & HNC & C$^{18}$O & C$_2$H & HCN & N$_2$H$^+$ & $^{13}$CO \\
    \hline % \\[0.2ex]
    CS & 1.00 & & & & & & & \\
    HCO$^+$ & 0.85 & 1.00 & & & & & & \\
    HNC & 0.90 & 0.92 & 1.00 & & & & & \\
    C$^{18}$O & 0.77 & 0.80 & 0.81 & 1.00 & & & & \\
    C$_2$H & 0.60 & 0.70 & 0.67 & 0.61 & 1.00 & & & \\
    HCN & 0.81 & 0.94 & 0.90 & 0.75 & 0.70 & 1.00 & & \\
    N$_2$H$^+$ & 0.80 & 0.76 & 0.82 & 0.73 & 0.56 & 0.71 & 1.00 & \\
    $^{13}$CO & 0.77 & 0.82 & 0.81 & 0.88 & 0.63 & 0.79 & 0.69 & 1.00 \\
    \hline
  \end{tabular}
\end{table*}
\begin{table*}
  \centering
  \caption{The eigenvectors and eigenvalues of the principal components (PC) derived from the
  correlation matrix listed in Table \ref{tab:cor_matrix}.}
  \label{tab:evectors_val}
  \begin{tabular}{c c c c c c c c c c}
    \hline% \\[0.2ex]
     & Percentage of variance & CS & HCO$^+$ & HNC & C$^{18}$O & C$_2$H & HCN
       & N$_2$H$^+$ & $^{13}$CO \\
    \hline % \\[0.2ex]
    PC 1 & 80.0 & 0.36 & 0.38 & 0.38 & 0.35 & 0.30 & 0.37 & 0.34 & 0.35 \\
    PC 2 & 6.3 & -0.25 & 0.05 & -0.10 & -0.17 & 0.86 & 0.15 & -0.36 & -0.06 \\
    PC 3 & 4.7 & -0.25 & -0.11 & -0.22 & 0.58 & -0.07 & -0.16 & -0.38 & 0.61 \\
    PC 4 & 3.8 & -0.05 & -0.33 & -0.17 & 0.21 & 0.38 & -0.50 & 0.64 & -0.03 \\
    PC 5 & 2.1 & 0.80 & -0.26 & 0.07 & -0.05 & 0.15 & -0.34 & -0.39 & 0.04 \\
    PC 6 & 1.4 & 0 & 0.03 & 0.13 & 0.68 & 0.01 & 0.02 & -0.18 & -0.70 \\
    PC 7 & 0.8 & -0.25 & -0.63 & 0.72 & -0.03 & 0.01 & 0.13 & -0.05 & 0.09 \\
    PC 8 & 0.7 & 0.22 & -0.52 & -0.48 & 0.10 & -0.01 & 0.66 & 0.11 & -0.03 \\
    \hline
  \end{tabular}
\end{table*}
\indent From the eigenvectors we constructed images of principal components (PC), by projecting the data set onto each of the eigenvectors. Shown in Figure \ref{fig:evec} are plots of eigenvectors of each molecule in the first four principal components. A (negative) positive value indicates the molecule is (anti-)correlated with others. The larger the value, the stronger the correlation. From the PC1 axis of Figure \ref{fig:evec}a it becomes clear that all of the eight molecules are positively correlated with each other. This can be visualised in Figure \ref{fig:pc1_4}a, which is the whole data set projected onto the first principal component only. This also resembles an `ideal' molecular distribution of the giant molecular cloud - the average intensity distribution of the species - and can be interpreted as the eight molecules being positively correlated on large scales. From the eigenvector plot of the second principal component (PC 2 axis of Figure \ref{fig:evec}a), C$_2$H stands out from other molecules; we suspect this is caused by scanning stripes resulting from the on-the-fly mapping procedure. This can be seen in the image of the second principal component (Figure \ref{fig:pc1_4}b); note the vertical and horizontal stripes of solid contours which resemble the scanning patterns presented in the C$_2$H integrated emission map (Figure \ref{fig:int2}d). In fact, the scanning stripes are present in all data taken simultaneously as listed in Table \ref{tab:mol_list} (the July, 2006 observation season). However, due to the lower signal-to-noise level of the C$_2$H emission, these scanning patterns become dominant in this map. We have explored the possibility of removing scanning artefacts by reconstructing the images without the second principal component. This indeed reduced the level of the artifacts, but it also subtracts a component of real emission from the image. Hence it provides a qualitative improvement, but the corrected image should not be used for quantitative studies.\\
\indent From the eigenvector plot of the third principal component (Figure \ref{fig:evec}b) we can see C$^{18}$O and $^{13}$CO are anti-correlated with other molecules. Spatially this anti-correlation appears as compact negative contours against a diffuse positive region in the third principal component image (Figure \ref{fig:pc1_4}c). This could be due to the differences between high and low density tracers discussed in Section \ref{sec:pca_density}, with the CO isotopologue distribution being more extended.\\
\indent The fourth principal component eigenvector plot (Figure \ref{fig:evec}c) shows an anti-correlation between N$_2$H$^+$ and HCO$^+$. The anti-correlation also appears clearly in the fourth principal component image (Figure \ref{fig:pc1_4}d), notably across  IRAS16172$-$5028 (shown by the arrow), possibly pointing to chemical differences across the object. Further discussion on this will be presented in Section \ref{sec:pca_N2Hp_HCOp}.\\
\indent The principal component analysis provides a way to quantify the spatial distribution of different molecules (with the constructed principal component images), and in placing different molecules into groups (with the eigenvectors). This results in minimising the number of different molecules required to characterise the basic properties of the cloud: CO isotopologues, HCO$^+$, N$_2$H$^+$, and C$_2$H. Implications will be discussed in the next section.\\

% ########################################################
%  Discussions
% ########################################################
\section{Discussion}\label{sec:discussion}
In this section we discuss the results from this multi-molecular line survey and the analysis techniques applied. We also focus on particularly interesting regions in the G333 giant molecular cloud.\\

% ########################################################
%   result from clump find
% ########################################################
\subsection{Size-line width, luminosity-size, and luminosity-line width correlation} \label{sec:larson}
\begin{figure*}
  \centering
  \includegraphics[]{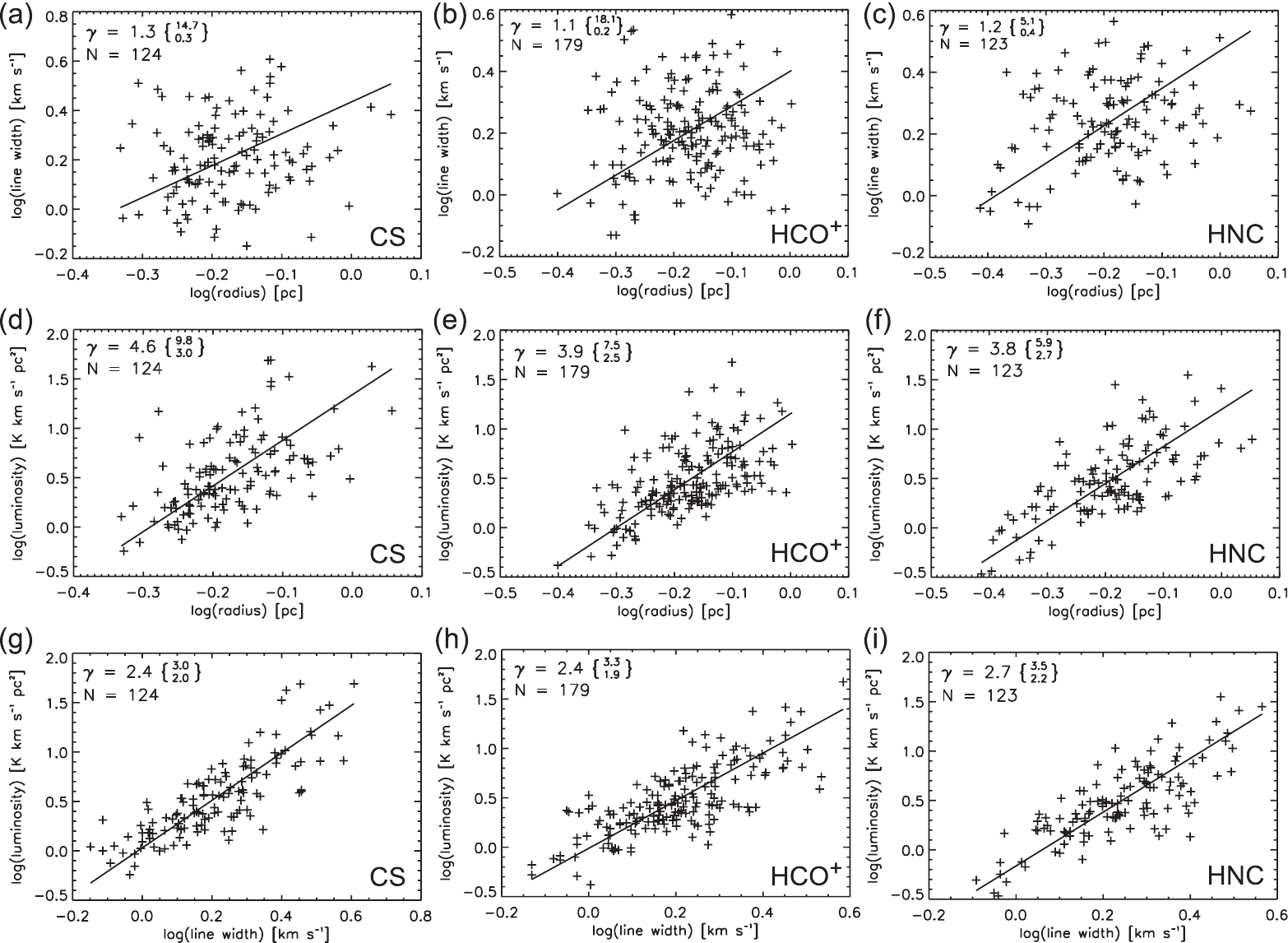}
  \caption[]{Plots of relationship between physical parameters for CS, HCO$^+$ and HNC
  clumps: (a) to (c) radius and line width; (d) to (f) luminosity and radius; (g) to (i)
  luminosity and line width.
  $\gamma$ indicates the slope of the ordinary least square (OLS) bisector fitted solid
  line, the values in curly brackets are the slope of the OLS ($X|Y$) and ($Y|X$).
  N is the number of clumps.
  Note that clumps with line widths smaller than 0.7 kms $^{-1}$ are discarded.}
  \label{fig:gcl_rel}
\end{figure*}
To examine the correlation among the clump properties of CS, HCO$^+$ and HNC, we have compared luminosities of the molecular line emission, line widths and radii in Figure \ref{fig:gcl_rel}. The C$_2$H clumps presented here were obtained analysing the main hyperfine component only. Hence the entire line luminosity has not been calculated. The luminosity was calculated with
\begin{equation}\label{equ:lum}
    L=(d[pc])^2(\frac{\pi}{180 \times 3600})^2 (r_x r_y) \int T {\rm d} v\, ,
\end{equation}
where $d$ is the distance to the GMC (3.6 kpc), $r_x$ and $r_y$ are the radii (arcsecond) of the two principal axes and $\int T {\rm d} v$ (K km s$^{-1}$) is the sum of emission at the maximum position. We have discarded clumps with line widths below 0.7 km s$^{-1}$, as this is less than twice the effective spectral resolution of 0.3 km s$^{-1}$ after hanning smoothing and binning the data. The slope of the ordinary least square (OLS) bisector \citep*{Feigelson1992} line of best fit ($\gamma$), estimates the structural relationship between variables $X$ and $Y$, without assuming whether $Y$ depends on $X$ or vice versa. Values inside the curly brackets are the slopes of OLS ($X|Y$), which minimises the residuals in $X$, and OLS ($Y|X$) which minimises the residuals in $Y$. The number of clumps ($N$) are shown on the top left corners of correlation plots. \\
\indent \citet*{Larson1981} empirically showed, utilising data from several molecular cloud surveys, that the line width $\Delta V$ appears to scale with cloud size $r$, $\Delta V \varpropto r^\gamma$. The common value quoted for $\gamma$ is $\sim 0.4 \pm 0.1$ \citep*[see][and references therein]{Mac_Low2004}. However, significant deviations have been reported \citep*[e.g.][]{{Loren1989a},{Caselli1995},{Plume1997}}. Consistent with our analysis of $^{13}$CO and C$^{18}$O line emission (WLB2008), we find no correlation between line width and radius for CS, HCO$^+$ and HNC clumps. The line width-size (note radius is given, not diameter) plots (Figure \ref{fig:gcl_rel}a-c) show a large scatter, hence the fitted slopes are poorly constrained. Note the large differences between OLS ($X|Y$) and ($Y|X$) slopes. Other line width-size relation studies \citep*[e.g.][ and references therein]{Schneider2004} also found no line width-size relation, and suggested its possible dependance on clump identification procedure.\\
\indent There is a clear correlation between luminosity and radius among CS, HCO$^+$ and HNC clumps. The fitted OLS bisector slopes $\gamma$ (3.8 to 4.6) are slightly steeper than found for $^{13}$CO and C$^{18}$O ($\sim3.0$). Since luminosity implicitly depends on $r^3$, one would expect $\gamma \thickapprox 3$, assuming a constant column density across the clumps. However, no statistical significance can be assigned to the value of slope, due to the inherent bias in clump finding algorithms, that tend to find larger clumps in denser regions \citep*{Schneider2004}.\\
\indent Similarly, luminosity and line width exhibits a strong correlation; this is not surprising as luminosity is calculated from the clump sizes and line widths. We do note that the $^{13}$CO and C$^{18}$O clumps from previous work (WLB2008) were found using a different clump-finding algorithm, {\sc cprops}.\\

% ########################################################
%   result from PCA
% ########################################################
\subsection{Physical and chemical properties from PCA}
In Section \ref{sec:PCA} we presented the results of principal component analysis, finding
correlations or anti-correlations between molecular species. In the following sections, we
discuss these in terms of the physical and chemical properties of the GMC.\\

\subsubsection{High and low density tracers} \label{sec:pca_density}
The eigenvector plot of the third principal component shows that $^{13}$CO and C$^{18}$O are anti-correlated with the other molecules (Figure \ref{fig:evec}b); this can be visualised spatially from the projected principal component image as shown in Figure \ref{fig:pc1_4}c. In this third PC image the compact dashed (negative) contours contrast spatially with the diffuse solid (positive) contours. This is due to C$^{18}$O and $^{13}$CO ($J=1\to0$) being low density tracers, and other molecules presented here being high density tracers (see Table \ref{tab:mol_list} column 5). $^{13}$CO and C$^{18}$O are present in both low and high density regions, whereas the other molecules are present in dense gas only. Hence in diffuse gas $^{13}$CO and C$^{18}$O are in `excess' compared to the high density tracers.\\

\subsubsection{The anti-correlation of N$_2$H$^+$ and HCO$^+$} \label{sec:pca_N2Hp_HCOp}
The eigenvector plot for the fourth principal component (Figure \ref{fig:evec}c) indicates the two ionic species N$_2$H$^+$ and HCO$^+$ are anti-correlated. According to models of dense clouds the major formation pathway of N$_2$H$^+$ is via ion-molecule reactions between N$_2$ and H$_3^+$ \citep*{Nejad1990}. It is mainly destroyed by reactions with CO \citep*{Tafalla2004} and by recombination with electrons in hot regions \citep*{Sternberg1995}. The following chemical equations summarise the mentioned reactions, 
\begin{eqnarray} \label{eqn:N2Hp_form}
	\mathrm{N_2} + \mathrm{H_3^+} \longrightarrow \mathrm{N_2H^+} + \mathrm{H_2}\, ,& 
\end{eqnarray}
\begin{eqnarray}\label{eqn:N2Hp_CO}
	\mathrm{N_2H^+} + \mathrm{CO} \longrightarrow \mathrm{HCO^+} + \mathrm{H_2}\, ,&
\end{eqnarray}
\begin{eqnarray}\label{eqn:N2Hp_e}
	\mathrm{N_2H^+} + e \longrightarrow \mathrm{N_2} + \mathrm{H}\, .
\end{eqnarray}
\noindent HCO$^+$ can be formed by ion-molecule reaction between N$_2$H$^+$ and CO (Equation \ref{eqn:N2Hp_CO}), or between CO and H$_3^+$ \citep*{Sternberg1995} as shown in Equation \ref{eqn:HCOp_CO} below, 
\begin{eqnarray}\label{eqn:HCOp_CO}
	\mathrm{CO} + \mathrm{H_3^+} \longrightarrow \mathrm{HCO^+} + \mathrm{H_2}\, .&
\end{eqnarray}
\noindent In cold dense cores, CO, the major destroyer of N$_2$H$^+$, will be depleted, so that the N$_2$H$^+$ abundance increases. With the lack of CO molecules, HCO$^+$ loses both of its major pathways (CO reacting with N$_2$H$^+$ or H$_3^+$), so the abundance of HCO$^+$ decreases. On the other hand, in warm regions where CO is not depleted, it reacts with N$_2$H$^+$ and produces HCO$^+$ causing a decrease in N$_2$H$^+$ and increase in HCO$^+$. CO reactions with H$_3^+$ further increase the HCO$^+$ abundance.\\
\indent Prominent sites to study this phenomenon are located in the vincinity of H{\sc ii} regions, i.e. in photodissociation or photon dominated regions (PDRs), where at the cloud surface a high HCO$^+$ and a low N$_2$H$^+$ abundance is expected. In contrast, in the inner part of the cloud, where material is collapsing to form a dense core, CO is depleted causing an increase in N$_2$H$^+$ and a drop in HCO$^+$ abundance. Thus HCO$^+$ and N$_2$H$^+$ should be anti-correlated. \\
\begin{figure}
  \centering
  \includegraphics[]{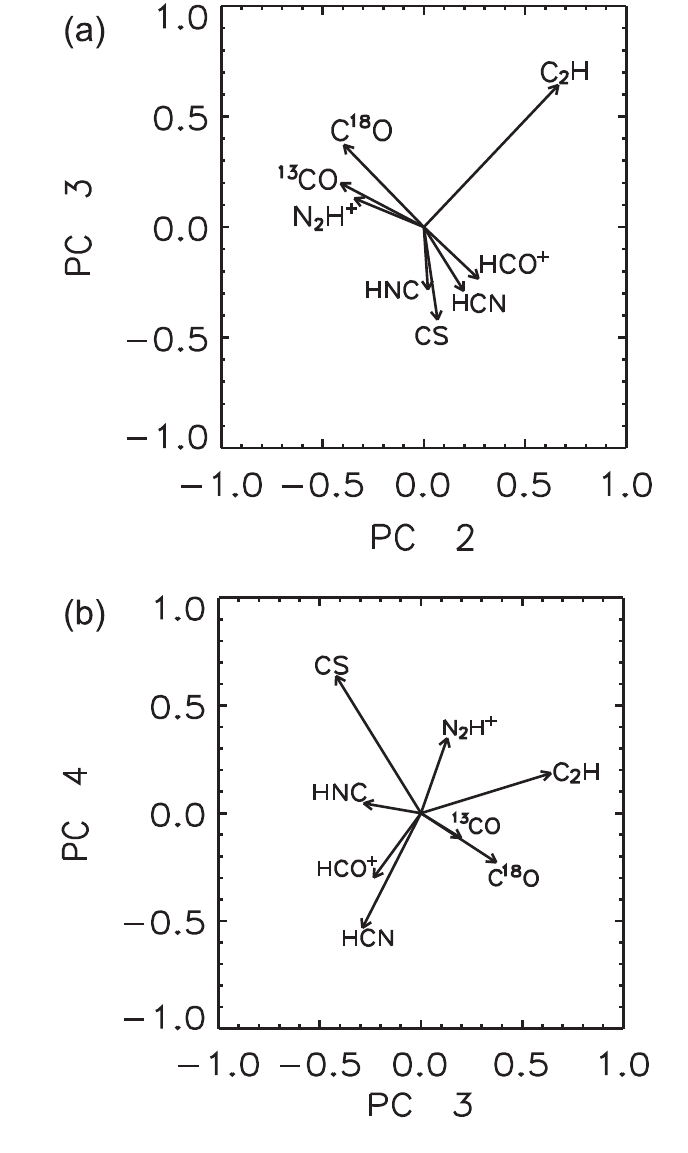}
  \caption[]{The eigenvectors of principal components two, three and four of the gas near the H{\sc ii}
  region associated with IRAS16172$-$5028.}
  \label{fig:evec_iras}
\end{figure}
\begin{figure}
  \centering
  \includegraphics[]{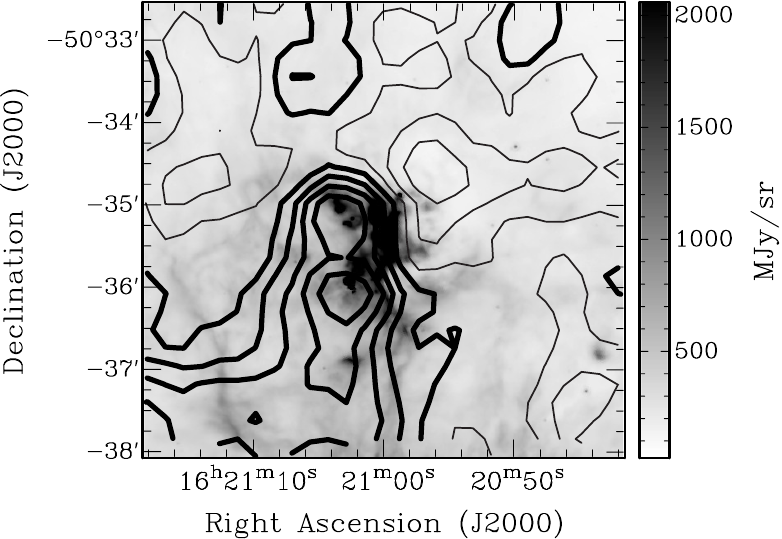}
  \caption[]{An image of the second principal component (contours) of the gas near the H{\sc ii} region associated
    with IRAS16172$-$5028 overlaid on the GLIMPSE 8.0-$\mu$m image (grey scale). The thin contours
    show regions with negative correlation to this component (mainly N$_2$H$^+$) and the thick 
    contours are regions with positive correlation (mainly HCO$^+$).}
  \label{fig:pc2_G4_iras}
\end{figure}
\paragraph{IRAS16172$-$5028}\label{par:iras1}
\indent The H{\sc ii} region associated with IRAS16172$-$5028 provides a good candidate to test whether N$_2$H$^+$ and HCO$^+$ are anti-correlated. As suggested by the compact intense N$_2$H$^+$ emission, this source has a cold high density component, but is immersed in a strong PDR according to the bright 8.0-$\mu$m emission. To exclude the effects of large-scale spatial contamination, we applied PCA to the integrated emission map of this region. Listed in Table \ref{tab:evectors_val_iras} are the eigenvalues and eigenvectors for this analysis; the eigenvectors of the second, third and fourth principal components are also plotted in Figure \ref{fig:evec_iras}; where the anti-correlation of HCO$^+$ and N$_2$H$^+$ is clearly seen. In the eigenvector plot of the second and third principal components (Figure \ref{fig:evec_iras}a), C$_2$H has the strongest positive correlation among the molecules. This is due to its spatially wide-spread emission towards the south-eastern region, compared to, for example, N$_2$H$^+$. Figure \ref{fig:pc2_G4_iras} shows the GLIMPSE 8.0-$\mu$m image (grey scale) overlaid with contours of the second principal component of the H{\sc ii} region associated with IRAS16172$-$5028, demonstrating that there are two regions with molecular emission that are anti-correlated. An examination of the eigenvectors shows that C$_2$H, HCO$^+$ (thick contours) and N$_2$H$^+$ (thin contours) contribute significantly to the variance in the second component and are anti-correlated, as expected for the hypothesis above. Note that the 8.0-$\mu$m emission (PAHs and warm gas) coincides well with the thick contours (HCO$^+$). \\
\begin{table*}
  \centering
  \caption{The eigenvectors and eigenvalues of the principal components (PC) of the gas near the
  H{\sc ii} region associated with IRAS16172$-$5028.}
  \label{tab:evectors_val_iras}
  \begin{tabular}{c c c c c c c c c c}
    \hline% \\[0.2ex]
     & Percentage of variance & CS & HCO$^+$ & HNC & C$^{18}$O & C$_2$H & HCN
       & N$_2$H$^+$ & $^{13}$CO \\
    \hline % \\[0.2ex]
    PC 1 & 84.8 & 0.36 & 0.37 & 0.37 & 0.35 & 0.31 & 0.36 & 0.35 & 0.35 \\
    PC 2 & 6.7 & 0.07 & 0.27 & 0.02 & -0.39 & 0.66 & 0.20 & -0.34 & -0.41 \\
    PC 3 & 3.4 & -0.42 & -0.23 & -0.28 & 0.37 & 0.64 & -0.29 & 0.13 & 0.20 \\
    PC 4 & 2.3 & 0.64 & -0.30 & 0.05 & -0.23 & 0.19 & -0.53 & 0.35 & -0.11 \\
    PC 5 & 1.4 & 0.35 & -0.07 & 0.05 & 0.20 & 0.04 & -0.21 & -0.78 & 0.43 \\
    PC 6 & 0.6 & 0.17 & -0.03 & 0.03 & 0.70 & -0.06 & -0.02 & -0.11 & -0.68 \\
    PC 7 & 0.4 & -0.21 & -0.63 & 0.72 & -0.06 & 0.10 & 0.18 & -0.08 & -0.04 \\
    PC 8 & 0.3 & -0.31 & 0.50 & 0.51 & 0 & -0.05 & -0.62 & -0.02 & -0.06 \\
    \hline
  \end{tabular}
\end{table*}
\begin{figure}
  \centering
  \includegraphics[]{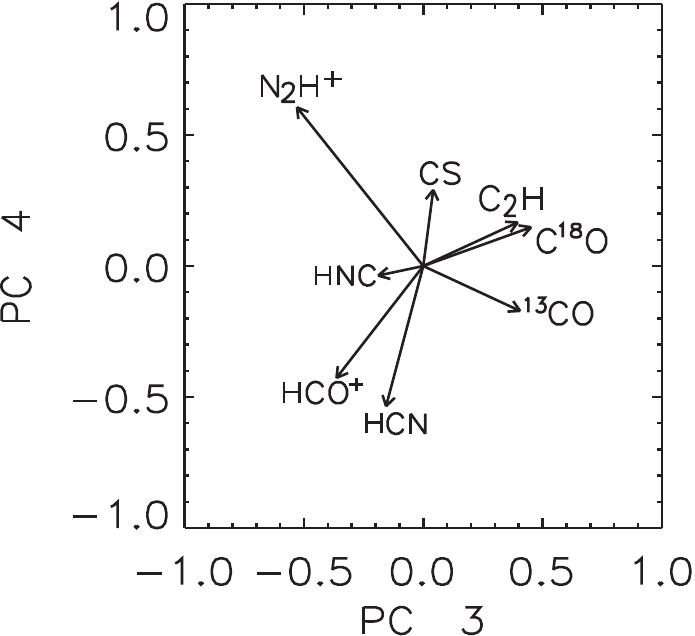}
  \caption[]{The eigenvectors of principal components two, three and four of the gas near the H{\sc ii}
  region associated with IRAS16177$-$5018.}
  \label{fig:evec_sh}
\end{figure}
\indent The second and third principal components of this region also show that $^{13}$CO and C$^{18}$O are correlated with N$_2$H$^+$ (Figure \ref{fig:evec_iras}a), which contradicts the idea suggesting that CO is the main destroyer of N$_2$H$^+$. However, we believe this can be explained by considering the different density conditions that these two species are tracing. CO is a low density tracer having a critical density of $\sim 10^3$ cm$^{-3}$, while N$_2$H$^+$ is a high density tracer with a critical density of $\sim 10^5$ cm$^{-3}$. Therefore CO is tracing both low and high density gas along the line of sight. Since maps of integrated emission were used for the PCA, the $^{13}$CO and C$^{18}$O may arise from a molecular envelope almost devoid of N$_2$H$^+$, while the bulk of the N$_2$H$^+$ emission is arising from the core.\\
\begin{table*}
  \centering
  \caption{The eigenvectors and eigenvalues of the principal components (PC) of the gas near the
  H{\sc ii} region associated with IRAS16177$-$5018.}
  \label{tab:evectors_val_sh}
  \begin{tabular}{c c c c c c c c c c}
    \hline% \\[0.2ex]
     & Percentage of variance & CS & HCO$^+$ & HNC & C$^{18}$O & C$_2$H & HCN
       & N$_2$H$^+$ & $^{13}$CO \\
    \hline % \\[0.2ex]
    PC 1 & 86.0 & 0.36 & 0.36 & 0.37 & 0.35 & 0.31 & 0.36 & 0.35 & 0.36 \\
    PC 2 & 5.4 & -0.01 & -0.08 & 0.08 & -0.46 & 0.79 & 0.18 & -0.09 & -0.34 \\
    PC 3 & 3.0 & 0.04 & -0.36 & -0.19 & 0.45 & 0.40 & -0.16 & -0.53 & 0.41 \\
    PC 4 & 2.8 & 0.29 & -0.43 & -0.04 & 0.15 & 0.17 & -0.54 & 0.61 & -0.17 \\
    PC 5 & 1.1 & -0.81 & 0.25 & -0.14 & 0.21 & 0.27 & -0.16 & 0.33 & 0.11 \\
    PC 6 & 0.7 & -0.30 & -0.70 & 0.24 & -0.02 & -0.14 & 0.52 & 0.22 & 0.16 \\
    PC 7 & 0.6 & -0.01 & 0 & -0.12 & 0.62 & -0.01 & 0.31 & -0.07 & -0.71 \\
    PC 8 & 0.4 & -0.18 & -0.01 & 0.85 & 0.12 & -0.04 & -0.36 & -0.26 & -0.17 \\
    \hline
  \end{tabular}
\end{table*}
\begin{figure}
  \centering
  \includegraphics[]{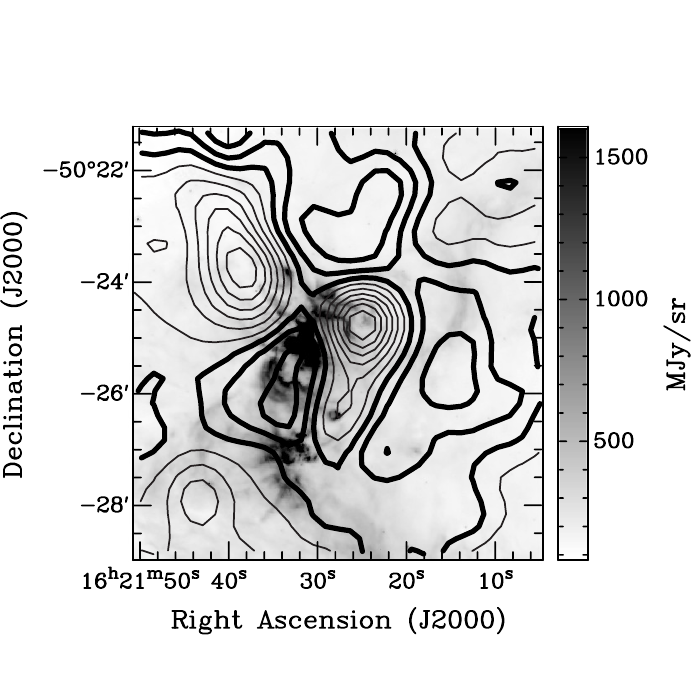}
  \caption[]{An image of the third principal component (contours) of the gas near the H{\sc ii} region associated
    with IRAS16177$-$5018 overlaid on the GLIMPSE 8.0-$\mu$m image (grey scale). The thin contours
    showing regions with negative correlations (mainly N$_2$H$^+$) and the thick 
    contours are regions with positive correlations (mainly $^{13}$CO and C$^{18}$O).}
  \label{fig:pc3_G4_sh}
\end{figure}
\begin{figure}
  \centering
  \includegraphics[]{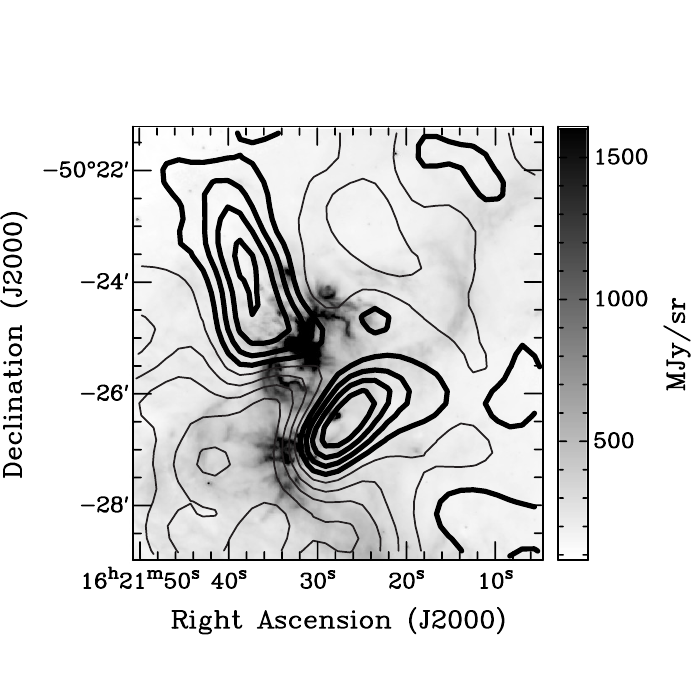}
  \caption[]{Similar to Figure \ref{fig:pc3_G4_sh} but the fourth principal component (contours) overlaid on the GLIMPSE 8.0-$\mu$m image (grey scale). The thin contours
    show regions with negative correlation to this component (mainly HCO$^+$) and the thick contours
    are regions with positive correlation (mainly N$_2$H$^+$).}
  \label{fig:pc4_G4_sh}
\end{figure}
\paragraph{IRAS16177$-$5018}\label{par:iras2}
Another example is the molecular emission associated with IRAS16177$-$5018 as shown in Figures \ref{fig:pc3_G4_sh} and \ref{fig:pc4_G4_sh}. The eigenvalues and eigenvectors of the principal components are listed in Table \ref{tab:evectors_val_sh}, while the eigenvectors of the third and fourth principal components are plotted in Figure \ref{fig:evec_sh}. Similar to the other source IRAS16172$-$5028, this region also shows an anti-correlation of N$_2$H$^+$ and HCO$^+$ in the fourth principal component (Figure \ref{fig:pc4_G4_sh}). On the other hand, shown in the third principal component (Figure \ref{fig:pc3_G4_sh}) is an anti-correlation of N$_2$H$^+$ and CO isotopologues.\\
\indent The PCA also indicates that HCN and HCO$^+$ are correlated, as shown in the eigenvector plots of the entire region (Figure \ref{fig:evec}). This correlation has also been suggested by other studies, such as by \cite*{Turner1977} for Orion-KL. IRAS16172$-$5028 and IRAS16177$-$5018 are both known to be associated with an H{\sc ii} region, contain cold molecular gas of high density (bright compact N$_2$H$^+$ emission), and have a low photon exposure. As expected by chemical models, both sources clearly show that N$_2$H$^+$ and HCO$^+$ are anti-correlated.\\

% ########################################################
\subsection{The infrared ring} \label{sec:IRring}
\begin{figure*}
  \centering
  \includegraphics[]{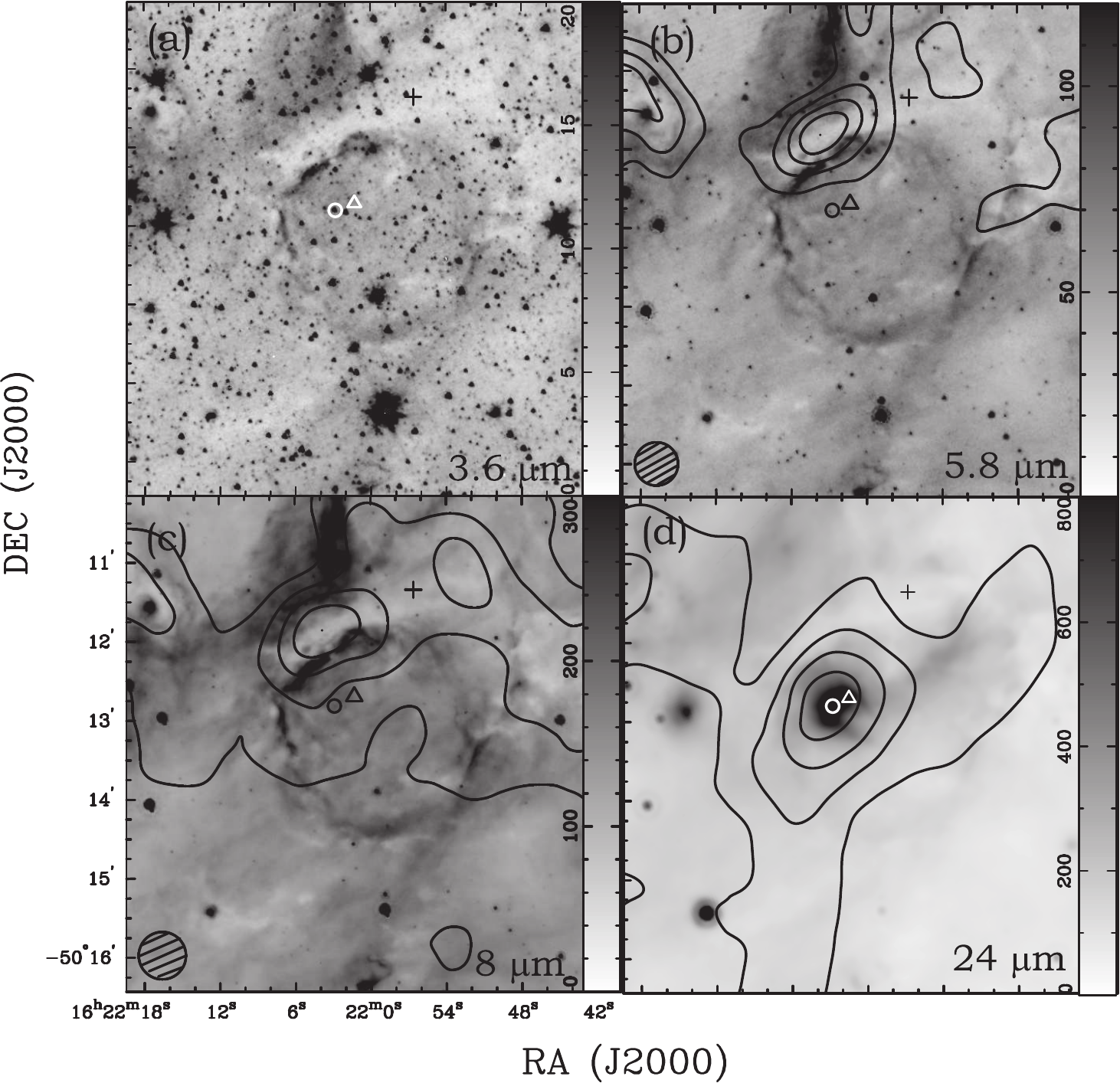}
  \caption[]{(a) GLIMPSE 3.6-$\mu$m image; (b) GLIMPSE 5.8-$\mu$m image overlaid with 1.2-mm dust continuum contours \citep{Mookerjea2004}; (c) GLIMPSE 8.0-$\mu$m images overlaid with N$_2$H$^+$ integrated emission contours and (d) MIPSGAL 24-$\mu$m image overlaid with MGPS 843-MHz radio continuum contours of the infrared ring discussed in Section \ref{sec:IRring}. The grey scale is stretched to 0 - 20 MJy sr$^{-1}$ for the GLIMPSE 3.6-$\mu$m, 0 - 120 MJy sr$^{-1}$ for the 5.8-$\mu$m image, 0 - 300 MJy sr$^{-1}$ for the 8-$\mu$m image and 0 - 800 MJy sr$^{-1}$ for the MIPSGAL image. The contours are at multiples of 20 per cent level of each of the emission peaks. The hatched circles represent the beam size of the respective contour maps. The small circle indicates the position of the {\it MSX} point source G333.5114-00.2798, the triangle marks the position of the {\it IRAS} source, IRAS16182$-$5005 (resolution of $\sim 4$ arcmin) and the cross shows the position of a far-infrared source \citep[resolution of $\sim 1$ arcmin]{Karnik2001} which peaks at 100-$\mu$m. Note the ring feature present at 5.8 and 8.0-$\mu$m, which is dominated by PAHs, the bright 24-$\mu$m emission which lies inside this ring. Both the 1.2-mm dust continuum and the integrated N$_2$H$^+$ emission coincide with the infrared dark filaments, which appear white in the 5.8 and 8.0-$\mu$m grey scale images.}
  \label{fig:IRring_4img_w_pntsrc}
\end{figure*}

Another note worthy feature in G333 is a ring of mid-infrared emission as shown in Figure \ref{fig:IRring_4img_w_pntsrc}, the northern portion of which is aligned with an arc of N$_2$H$^+$ emission (see the top arrow on Figure \ref{fig:int2}). This ring feature is visible in the GLIMPSE 5.8 and 8.0-$\mu$m (Figure \ref{fig:IRring_4img_w_pntsrc} top right and bottom left panels) images, but is marginally detectable in 3.6- (Figure \ref{fig:IRring_4img_w_pntsrc} top left panel) and 4.5-$\mu$m images. The GLIMPSE 5.8 and 8.0-$\mu$m channels of the {\it Spitzer} IRAC instrument are dominated by PAH emission excited by nearby ultraviolet sources \citep{Reach2006}. Lying inside the emission ring but offset towards the edge, is bright MIPSGAL \citep{Carey2006} 24-$\mu$m emission (Figure \ref{fig:IRring_4img_w_pntsrc} bottom right panel), spatially coincident with the {\it MSX} point source G333.5114$-$00.2798 \citep[marked with a white circle;][]{Egan2003} and an {\it IRAS} source IRAS16182$-$5005 (marked with triangle). The absence of the emission ring at 24-$\mu$m further suggests that the emission originates from PAHs, and that there is a lack of thermal dust emission.\\
\indent Comparing the MIPSGAL 24-$\mu$m emission with the 843-MHz radio continuum (Figure \ref{fig:IRring_4img_w_pntsrc}d contours) from the Molonglo Galactic Plane Survey \citep[MGPS,][]{Green1999}, it is clear that the radio continuum is associated with the 24-$\mu$m emission. A search of the SIMBAD astronomical database and VizieR catalogue service\footnote{http://cdsweb.u-strasbg.fr/} did not lead to the identification of any supernova or X-ray source, suggesting there is an H{\sc ii} region inside the 8-$\mu$m emission ring. Shown in Figure \ref{fig:IRring_4img_w_pntsrc}b is the 5.6-$\mu$m emission ring (grey scale) overlaid with 1.2-mm dust continuum contours \citep{Mookerjea2004}; the 1.2-mm dust continuum lies next to the bright 8-$\mu$m rim and is coincident with a dark filament (appearing white in the grey scale image). According to \citet{Mookerjea2004}, this source (MMS16) has a density of $\sim2 \times 10^4$ cm$^{-3}$. The 1.2-mm dust continuum emission agrees well with the arc of N$_2$H$^+$ emission (Figure \ref{fig:IRring_4img_w_pntsrc}c) described above. There is no detectable dust continuum from the south-western region of the ring. Molecular emission is detected from this part but it is weak and diffuse, indicating a relative low density region compared to the dust. Given the completeness and symmetry of the ring, one might  expect the H{\sc ii} region to lie near the centre. Since it does not, we suggest the possibility that this is due to density differences in the gas surrounding the infrared ring. In general H{\sc ii} regions are density bounded, and the pressurised H{\sc ii} gas breaks out of the cloud into lower-density gas, creating a champagne flow \citep{Stahler2005}. The inhomogeneous density (from our molecular line data) allows the ionised gas to spread out further towards the south-west compared to the north-east rim, creating a near circular 8-$\mu$m emission structure with the driving source lying on the edge. However the extreme symmetry also suggests the ring could be a pre-existing structure that is currently being illuminated by the H{\sc ii} region. In either way this infrared emission ring must be related to the H{\sc ii} region.\\
\indent Another notable feature is the spatial coincidence of mm dust continuum, N$_2$H$^+$ emission and the infrared dark filament (which appears as white in the grey scale image) to the north of the ring (Figure \ref{fig:IRring_4img_w_pntsrc}b and c). As CO is frozen on to grains in dense cold gas \citep[see e.g.][]{{Kramer1999},{Bacmann2002}}, causing it to be depleted, and the ion-molecule formation reaction for N$_2$H$^+$ proceeds regardless of temperature, N$_2$H$^+$ is a good candidate for tracing cold dense gas where CO is depleted. With mm dust continuum tracing cold dust along with N$_2$H$^+$ tracing cold dense gas, it is not surprising that they coincide with the infrared dark filament.
%
% #############################################
% Summary
% #############################################
\section{Summary}\label{sec:summary}
In this paper we have presented data from a multi-molecular line survey of the southern star forming region, the G333 giant molecular cloud complex. For this survey, we have exploited the Mopra MMIC receiver and the 8-GHz bandwidth UNSW-Mopra Spectrometer, resulting in over twenty multi-molecular transition maps, with velocity resolution of $\sim0.1$ km s$^{-1}$. We have presented total intensity maps of molecules with bright emission and have also discussed the velocity structure of the G333 molecular cloud. To further characterise the physical and chemical properties, we have carried out common analysis techniques such as {\sc gaussclumps} to obtain distributions of CS, HCO$^+$, HNC and C$_2$H emission. We have also performed principal component analysis on the data set, to visualise and parameterise the differences between the spatial distribution of molecules. In this work, we have found:
\begin{enumerate}
  \item Differences in spatial and velocity distribution among different molecules. We found that the spatial distribution of CS, HCO$^+$, HCN and HNC are similar on large scales, while N$_2$H$^+$ seems to trace preferentially the very densest regions. C$_2$H is only detected close to bright infrared emission regions. The detected molecules all have similar velocity distributions.
  \item The velocity gradient across the GMC complex noted in $^{13}$CO (BWC2006) and C$^{18}$O (WLB2008) is also present in CS, HCO$^+$ and HNC.
  \item CS, HCO$^+$ and HNC emission maps were decomposed with {\sc gaussclumps} in three dimensions. We found no correlation between clump radius and line width, but a clear correlation between luminosity and radius. Accounting for saturation effects in the CS ($J=2 \to 1$) and HCO$^+$ ($J=1 \to 0$) lines toward clumps, we obtain column densities of $\sim 10^{12}$ to $\sim 10^{14}$ cm$^{-2}$.
  \item An alternative approach used to characterise this data set was principal component analysis (PCA). PCA separates molecules into low ($^{13}$CO and C$^{18}$O) and high (the rest) density tracers, identifies anti-correlations between HCO$^+$ and N$_2$H$^+$, correlations between HCN and HCO$^+$, and helps to explore scanning patterns of the `on-the-fly' mapping.
  \item A noteworthy `ring-like' structure of the GMC is present in the GLIMPSE 8-$\mu$m image, associated with an H{\sc ii} region as suggested by radio continuum emission inside the ring. The molecular line data (especially N$_2$H$^+$) shows gas being swept up and compressed by the ring.
\end{enumerate}

\section*{Acknowledgments}
The Mopra Telescope is part of the Australia Telescope and is funded by the Commonwealth of Australia for operation as National Facility managed by CSIRO. The University of New South Wales Mopra Spectrometer Digital Filter Bank used for the observations with the Mopra Telescope was provided with support from the Australian Research Council, together with the University of New South Wales, University of Sydney and Monash University. PAJ acknowledges partial support from Centro de Astrof\'\i sica FONDAP 15010003 and the GEMINI-CONICYT FUND. This research (GLIMPSE and MIPSGAL images) has made use of the NASA/IPAC Infrared Science Archive which is operated by the Jet Propulsion Laboratory, California Institute of Technology, under contract with NASA.

% ------------------------------------------------------------------------------
% Bibliography
% ------------------------------------------------------------------------------
%\addcontentsline{toc}{chapter}{References}
%\renewcommand{\bibname}{References}
%\bibliographystyle{unsrtnat}
\bibliographystyle{mn2e}
\bibliography{G333_paper3}

% ------------------------------------------------------------------------------
% Appendix
% ------------------------------------------------------------------------------
%\appendix
%\section{appendix section}

\bsp

\label{lastpage}

\end{document}